\documentclass[aps,twocolumn,amsmath,amssymb,nofootinbib,superscriptaddress]{revtex4-2}
\usepackage{times}
\usepackage[pdftex]{graphicx}
\usepackage{dcolumn}
\usepackage{bm}
\usepackage{amsmath}
\usepackage{amsthm}
\usepackage{braket}
\usepackage{indentfirst}
\usepackage[colorlinks]{hyperref}
\usepackage[dvipsnames]{xcolor}
\usepackage{xcolor}
\usepackage{mathtools}
\DeclarePairedDelimiter{\ceil}{\lceil}{\rceil}

\definecolor{aqua}{RGB}{69,139,116}

\usepackage{subfigure}
\usepackage{algorithm}
\usepackage{algorithmicx}
\usepackage{algpseudocode}
\usepackage{amsmath}
\usepackage{verbatim}
\usepackage{makecell}
\usepackage{lettrine}
\usepackage[normalem]{ulem}

\begin{document}
\title{Circuit-Noise-Resilient Virtual Distillation}
\author{Xiao-Yue Xu}
\affiliation{Henan Key Laboratory of Quantum Information and Cryptography, Zhengzhou, Henan 450000, China}
\author{Chen Ding}
\affiliation{Henan Key Laboratory of Quantum Information and Cryptography, Zhengzhou, Henan 450000, China}
\author{Shuo Zhang}
\affiliation{Henan Key Laboratory of Quantum Information and Cryptography, Zhengzhou, Henan 450000, China}
\author{Wan-Su Bao}
\email{bws@qiclab.cn}
\affiliation{Henan Key Laboratory of Quantum Information and Cryptography, Zhengzhou, Henan 450000, China}
\affiliation{Hefei National Laboratory, University of Science and Technology of China, Hefei 230088, China}
\author{He-Liang Huang}
\email{quanhhl@ustc.edu.cn}
\affiliation{Henan Key Laboratory of Quantum Information and Cryptography, Zhengzhou, Henan 450000, China}
\affiliation{Hefei National Laboratory, University of Science and Technology of China, Hefei 230088, China}
\date{\today}

\begin{abstract}
\textbf{Quantum error mitigation (QEM) is vital for improving quantum algorithms' accuracy on noisy near-term devices. A typical QEM method, called Virtual Distillation (VD), can suffer from imperfect implementation, potentially leading to worse outcomes than without mitigation. To address this, we introduce Circuit-Noise-Resilient Virtual Distillation (CNR-VD), which includes a calibration process using simple input states to enhance VD's performance despite circuit noise, aiming to recover the results of an ideally conducted VD circuit. Simulations show that CNR-VD significantly mitigates noise-induced errors in VD circuits, boosting accuracy by up to tenfold over standard VD. It provides positive error mitigation even under high noise, where standard VD fails. Furthermore, our estimator's versatility extends its utility beyond VD, enhancing outcomes in general Hadamard-Test circuits. The proposed CNR-VD significantly enhances the noise-resilience of VD, and thus is anticipated to elevate the performance of quantum algorithm implementations on near-term quantum devices.}

\end{abstract} 

\maketitle

\noindent\textbf{\large{Introduction}}

In the pursuit of large-scale fault-tolerant quantum computers, the presence of noise arising from environmental noise and imperfect hardware remains a significant challenge.~\cite{Preskill2018quantumcomputing,Arute2019supremacy,huang2020superconducting,DingEvaluating2022}. The field of 
Quantum error mitigation (QEM) techniques~\cite{Endo2021Hybrid,Kandala2019Errormitigation,Li2017Efficient,Huang2023NearTerm,Qin2023ErrorStatistics,Yoshioka2022Generalized,McArdle2019errormitigated,Cai2021multi,ding2023noise} braves this challenge, offering hopeful solutions that reduce the impact of noise even in the absence of full-blown quantum error correction~\cite{Chiaverini2004Realization,Knill2000Theory,Cory1998Experimental,Terhal2015memories}.
Among these techniques, one typical QEM method called Virtual Distillation (VD)~\cite{Huggins2021Virtual,Koczor2021Exponential,Alireza2022Shadow,Huo2022Dual,Cai2021Resource,Xiong2022Permutation,Hong2022Logical,Siah2023dilution} has gained considerable attention. For a given $N$-qubit noisy quantum state $\rho$ and an observable $O$, the $n$-order VD algorithm aims to extract the ideal component from $\rho$ using $n$ copies of $\rho$. VD offers distinct advantages for exponential suppression of errors in expectation value estimation for observables, and its state-noise-agnostic nature.

However, employing VD algorithm for high-accuracy expectation values encounters two primary challenges. 
The first challenge is the coherent dismatch~\cite{Koczor2021Thedominant,Huggins2021Virtual,Koczor2021Exponential}, which can be mitigated through combinations of other QEM methods such as quantum subspace expansion~\cite{McArdle2019errormitigated,Colless2018Computation,li2017hybrid} in the state preparation circuit of $\rho$~\cite{Yoshioka2022Generalized}, or polynomial fitting of multiple orders of VD~\cite{Koczor2021Exponential}. 
The second challenge arises from noise in the VD circuit. The Hadamard-Test-like VD circuit~\cite{Koczor2021Exponential} utilized to efficiently implement VD consists of controlled derangement operator $D_n$ and controlled operator $O$. As the number of multi-qubit gates in the VD circuit increases linearly with $nN$~\cite{Koczor2021Exponential}, this can introduce non-negligible deviations in the estimated expectation values, potentially surpassing the initial errors in state preparation. Mitigating this error often involves treating the VD circuit as a part of the target circuit and leveraging additional QEM techniques~\cite{Koczor2021Exponential}, or employing shadow estimations~\cite{Huang2020Predicting,Huang2021Efficient,Elben2020Mixed} to avoid executing the VD circuit, albeit with an exponential measurement overhead~\cite{Alireza2022Shadow,Hong2022Logical}.

In this work, we focus on mitigating the noise in the VD circuit and propose the Circuit-Noise-Resilient Virtual Distillation (CNR-VD). In our method, we integrate a calibration procedure that utilizes the noisy VD circuit. During this procedure, the noisy state $\rho$ undergoes a transformation into easily-prepared states that are independent of $\rho$. The expectation values of these states are obtained via noisy VD (VD with a noisy VD circuit), and are subsequently employed to enhance mitigation outcomes, which is facilitated by a function derived from noisy VD. Numerical results confirm the effectiveness of CNR-VD, demonstrating a significant improvement in accuracy---up to an order of magnitude---compared to noisy VD. We also observe  that CNR-VD raises the gate noise threshold for VD, resulting in effective mitigation outcomes. Moreover, we compare the accuracy and efficiency of CNR-VD with two common VD-based methods designed to counter noise in the VD circuit, and observe the superiority of CNR-VD. 

\noindent\textbf{\large{Results}}
\begin{figure*}[t]
    \centering
    {\includegraphics[width=0.95\textwidth]{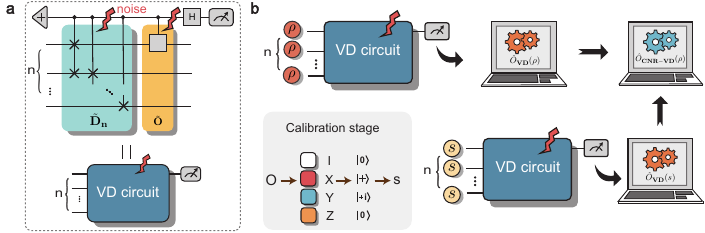}}    
    \caption{{\textbf{The noisy Virtual Distillation (VD) circuit and the circuit-noise-resilient virtual distillation (CNR-VD).} \textbf{a} The noisy VD circuit of $n$-order VD to estimate the expectation value of Pauli observable $O$. \textbf{b} The illustration of CNR-VD. In the $n$-order noisy VD (VD with noisy VD circuit), we use $n$ copies of the noisy state $\rho$ to compute the estimator $\hat{O}_\text{VD}(\rho)$. The CNR-VD estimator $\hat{O}_\text{CNR-VD}(\rho)$ integrates a calibration procedure where the noisy state $\rho$ undergoes a transformation into an easily-prepared state $s$. The expectation value $\hat{O}_\text{VD}(s)$ is obtained via noisy VD and subsequently employed to enhance mitigation outcomes. In the calibration procedure highlighted in the gray box in (b), we show one possible choice for $s$ which is the eigenstate of Pauli $O$ with eigenvalue $+1$.}}
    \label{fig1}
  \end{figure*}

\noindent\textbf{Virtual distillation with noisy VD circuit}

Suppose we have an $N$-qubit noisy quantum state $\rho$ which can be expressed as
\begin{equation}
  \rho=(1-\epsilon)|\psi\rangle\langle\psi|+\epsilon\sum_k\lambda_k|\psi_k\rangle\langle\psi_k|
\end{equation}
through spectral decomposition where $\epsilon \geq 0$ is the error rate, along with the ideal state $|\psi\rangle\langle\psi|$ and the noisy components $\sum_k\lambda_k|\psi_k\rangle\langle\psi_k|$ (satisfying $\sum_k\lambda_k=1$) of $\rho$. The $n$-order VD algorithm produces mitigated expectation value of observable $O$ as
\begin{equation}
    \langle O\rangle_\text{VD}=\frac{\text{Tr}[\rho^nO]}{\text{Tr}[\rho^n]}\approx\langle\psi|O|\psi\rangle+\mathcal{O}(\epsilon^n),
\end{equation}
where the error term $\mathcal{O}(\epsilon^n)$ is exponentially suppressed with $n$.
By means of the VD circuit, we do not require the physical preparation of $\rho^n$. The VD circuit consists of the controlled derangement operator $D_n$ containing $nN$ CSWAP gates and the controlled operator $O$.
Accordingly, the mitigated value is
\begin{equation}
  \langle O\rangle_\text{VD}=\frac{\text{Tr}[\rho^nO]}{\text{Tr}[\rho^n]}=\frac{\text{Tr}[D_n((\rho O)\otimes \rho^{\otimes n-1})]}{\text{Tr}[D_n\rho^{\otimes n}]},
\end{equation}
where Tr$[\rho^nO]$ is calculated by $2p_0(\rho,O)-1$ with the probability $p_0(\rho,O)$ of measuring the ancilla qubit $Q_0$ in the zero state. We can use the same method to estimate Tr$[\rho^n]$ by replacing $O$ with $I$ in the VD circuit. 

Now we will analyse the effects of the noise in the VD circuit. 
We consider the $k$-weight ($k\leq N$) Pauli observable $O=\prod_{\omega=1}^kO^{i_\omega}$ where the $\omega$-th non-trivial Pauli term on qubit $Q_{i_\omega}$ is $O^{i_\omega}\in\{\sigma_x,\sigma_y,\sigma_z\}$. Then the controlled operator $O$ is comprised of $k$ two-qubit controlled Pauli gates. For the gate noise assumption, we primarily focus on the $m$-qubit ($m\geq 1$) stochastic Pauli gate noise $\mathcal{E}^\text{P}_m$ with error rate $0<p_m<1$,
\begin{equation}
  \mathcal{E}^\text{P}_m(\rho)=(1-p_m)\rho+p_m\sum_{s=1}^{4^m-1}\delta_s^mP_s^m\rho P_s^m,
\end{equation}
where $P_s^m$ is the $s$-th Pauli operator on $m$ qubits and the coefficients $\{\delta_s^m\}_s$ satisfy $\delta_s^m\geq 0, \sum_s\delta_s^m=1$. 
Considering that measurement error mitigation protocols~\cite{Maciejewski2021modelingmitigation,Kwon2021Ahybrid,Chen2019Detector,Bravyi2021Mitigate,Maciejewski2020mitigationofreadout,Nation2021Scalable,Alistair2021Qubit} such as Tensor Product Noise (TPN)~\cite{geller2020rigorous,Bravyi2021Mitigate} can effectively suppress single-qubit readout errors in VD, which purely involves measurements on one qubit, we set up clean readout of the ancilla. We denote $\tilde{p}_0(\rho,O)$ to be the noisy version of $p_0(\rho,O)$ which is contaminated by the noise in the VD circuit, and we will pay our attention to the second-order VD. Then the outcome gained through noisy VD circuit can be expressed as a function 

\begin{equation}\label{eq-noise}
  \begin{aligned}
    2\tilde{p}_0(\rho,O)-1=\sum_{j=0}^{\ceil{\frac{k}{2}}}\sum_{t=0}^{\binom{k}{j}-1}a^k_{j,t}(p_1,p_2,p_3,O)A_j(t),
  \end{aligned}
\end{equation}
where $j$ denotes the number of unwanted SWAP gates generated by $\sigma_x/\sigma_y$ Pauli errors occurring on the controlled qubit of CSWAP gates, and $t$ represents the $t$-th combination of positions for these $j$ SWAPs. And $a^k_{j,t}(p_1,p_2,p_3,O)$ refers to the $\rho$-independent parameters for $A_j(t)$ (we define $A_0(0)=A_k(0)=\text{Tr}[\rho^2O]$) which can be written as 
\begin{equation}\label{eq-Aj}
  \begin{aligned}
    A_j(t)&=(1-\epsilon)^2\text{Tr}[O_1^{j,t}|\psi\rangle\langle\psi|]\cdot\text{Tr}[O_2^{j,t}|\psi\rangle\langle\psi|]\\&+\epsilon^2\sum_k\lambda_k^2\text{Tr}[O_1^{j,t}|\psi_k\rangle\langle\psi_k|]\cdot\text{Tr}[O_2^{j,t}|\psi_k\rangle\langle\psi_k|].
  \end{aligned}
\end{equation}
$A_j(t)$ is a combination of expectation values of observables $O_1^{j,t}$ and $O_2^{j,t}$ which meets $O=O_1^{j,t}+O_2^{j,t}$, and it derives from  the $t$-th combination of $j$ undesired SWAPs (See more details in the Supplementary Note 2).

\noindent\textbf{Circuit-Noise-Resilient virtual distillation}

We introduce the CNR-VD, which aims to extract Tr$[\rho^2O]$ attained by ideal VD (VD with noiseless VD circuit) based on the function in Eq.~\ref{eq-noise}. We begin with the simplest case involving a one-weight Pauli observable $O$ (equally applicable to $I$), which holds $A_j(t)=\text{Tr}[\rho^2O]$ for $j=0,1$ and $t=0$. Then Eq.~\ref{eq-noise} reduces to
\begin{equation}\label{eq-1weight}
  2\tilde{p}_0(\rho,O)-1=a(p_1,p_2,p_3,O)\text{Tr}[\rho^2O].
\end{equation}
The $\rho$-independent parameter $a(p_1,p_2,p_3,O)$ is amenable to calibration if we assume the temporal stability of Pauli error rate $p_i$($i=1,2,3$). In order to obtain these parameters, we add a calibration procedure using the noisy VD circuit where the noisy state $\rho$ is replaced by some separate state $s$ which can be easily prepared via single-qubit gates, as shown in the Fig.~\ref{fig1} \textbf{b}. The state $s$ can be any eigenstate of $O$ with eigenvalue $+1$ which satisfys Tr$[s^2O]=1$, and the parameter $a(p_1,p_2,p_3,O)$ in Eq.~\ref{eq-1weight} can be estimated by $2\tilde{p}_0(s,O)-1$ with an estimation error $\mathcal{O}(p_1)$ due to the noisy preparation of $s$. We can also attain the estimation of parameter $a(p_1,p_2,p_3,I)$ for Tr$[\rho^2]$ through noisy VD circuit with $O=I$.  
Then with the noisy VD estimator defined as $\hat{O}_\text{VD}(\rho)=(2\tilde{p}_0(\rho,O)-1)/(2\tilde{p}_0(\rho,I)-1)$,
we can establish the CNR-VD estimator  
\begin{equation}\label{eq-est1}
  \hat{O}_\text{CNR-VD}(\rho)=\frac{\left(\frac{2\tilde{p}_0(\rho,O)-1}{2\tilde{p}_0(s,O)-1}\right)}{\left(\frac{2\tilde{p}_0(\rho,I)-1}{2\tilde{p}_0(s,I)-1}\right)}=\frac{\hat{O}_\text{VD}(\rho)}{\hat{O}_\text{VD}(s)},
\end{equation}
which exhibits resilience to the noise in the VD circuit for any one-weight Pauli observable $O$. We can also employ the CNR-VD estimator to obtain reliable expectation values for higher-weight Pauli observables under moderate noise level of Pauli gate noise, with an estimation error $\mathcal{O}(p_1,p_2,p_3)$ as we ignore terms in Eq.~\ref{eq-noise} with $j>1$. When the assumption of stochastic Pauli gate noise model in our work dose not suit the practical scenario, the employment of randomized compiling~\cite{Wallman2016Noisetailoring,Erhard2019Characterizing,Hashim2021Randomizedcompiling} (RC) technique can enable us come close to the stochastic Pauli gate noise. Moreover, since the independence on $\rho$ of the parameters in Eq.~\ref{eq-1weight}, estimations of multiple noisy states can be carried out once the calibration procedure is executed to minimize circuit resource consumption, as long as the temporal stablility of gate noise holds at least during a certain time period.  

\begin{figure}[t]
  \centering
  {\includegraphics[width=0.48\textwidth]{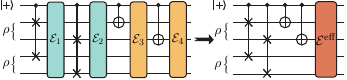}}    
  \caption{{\textbf{The effective noise of the second-order VD circuit for the two-qubit noisy state $\rho$ and two-weight Pauli observable $\sigma_x^{\otimes 2}$.} We define the general global Markovian noise of the $i$-th layer as $\mathcal{E}_{i+1}$. After relocating all the noise to the end of the VD circuit, we define the mixture of these noise conjugated by the layers behind as the effective noise $\mathcal{E}^\text{eff}$.}} 
  \label{fig2}
\end{figure}

Besides the local Pauli gate noise model, our CNR-VD estimator is also applicable to the general Markovian noise model without restrictions on locality. For the $n$-order VD, suppose the VD circuit has $l$ layers of noisy gates $\tilde{\mathcal{U}}_{1:l}=\prod_{j=1}^l\mathcal{E}^j\mathcal{U}_j$, where $\mathcal{E}^j$ is the Markovian noise acting on the $j$-th layer. After relocating all the noise to the end of the VD circuit, we define the mixture of these noise conjugated by the layers behind as the effective noise $\mathcal{E}^{\text{eff}}(\cdot)=\sum_iK_i\cdot K_i^\dagger$ of the VD circuit, where $K_i$ is the Kraus operator of the effective noise.  
Hence, the noisy VD circuit can be divided into an ideal VD circuit followed by the effective noise, as shown in Fig~\ref{fig2}. We define an $(nN+1)$-qubit state $\rho^\text{RD}=\rho^\text{R}\otimes\rho^\text{D}$ where $\rho^\text{R}$ is a single-qubit pure state and $\rho^\text{D}$ is an $(nN)$-qubit operator.
Then if the effective noise $\mathcal{E}^\text{eff}$ satisfies any of these two conditions:
\begin{itemize}
  \item There exists a set of Kraus operators such that: each Kraus operator $K_i=K_i^\text{R}\otimes K_i^\text{D}$ of $\mathcal{E}^\text{eff}$ is separate into subsystems $\rho^\text{R}$ and $\rho^\text{D}$, along with unitary $K_i^\text{R}$ and $K_i^\text{D}$.
  \item $\mathcal{E}^\text{eff}=\mathcal{E}^\text{eff,R}\otimes\mathcal{E}^\text{eff,D}$, along with the single-qubit unital noise channel $\mathcal{E}^\text{eff,R}$ and the CPTP noise channel $\mathcal{E}^\text{eff,D}$,
\end{itemize}
the CNR-VD estimator $\hat{O}_\text{CNR-VD}(\rho)$ can reduce the error in the expectation value indued by $\mathcal{E}^\text{eff}$ when the state preparation error for $s$ is negligible compared to $\mathcal{E}^\text{eff}$ (See proofs in the Supplementary Note 3). Since the depolarizing gate noise statisfys Condition 2 as it commutes with the CSWAP gates in the VD circuit, $\hat{O}_\text{CNR-VD}(\rho)$ show high tolerance to low-error-rate Pauli gate noise which is close to the depolarizing gate noise. We also establish other form of the CNR-VD estimator with loosen requirement of $\mathcal{E}^\text{eff}$ in the Supplementary Note 3.

Furthermore, we can extend our CNR-VD to encompass a broader range of quantum computing applications beyond QEM algorithms. By generalizing the operators $D_n$ and $O$ in the VD circuit to arbitrary Hermitian operator $U$, we can get a typical Hadamard-Test circuit. Then we can use the CNR-VD estimator to calculate the expectation value of $\text{Tr}[\tilde{\rho}U]$ and reduce the additional noise caused by the controlled-$U$. As for the input state $s$ used in the calibration procedure, we can choose any state that satisfys $\text{Tr}[sU]=1$ (See more details and simulations in the Supplementary Note 4).  
\begin{figure}[t]
  \centering
  {\includegraphics[width=0.48\textwidth]{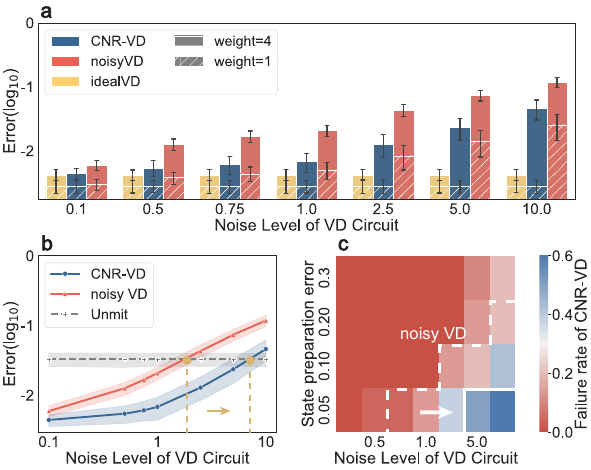}}    
  \caption{{\textbf{The accuracy of CNR-VD under Pauli gate noise with random states.} 
  \textbf{a} Results of 4-qubit random states under state preparation error rate $\epsilon=0.10$ obtained from CNR-VD, ideal VD (VD with noiseless VD circuit) and noisy VD (VD with noisy VD circuit) with increasing noise levels of the VD circuit and weights of observables. \textbf{b} Accuracy of 4-qubit random states under state preparation error rate $\epsilon=0.10$ shifting with an increase in the noise level. Yellow dots mark inflection points, dashed lines show noise thresholds, and grey dashed lines denote unmitigated estimation outcomes (Unmit). \textbf{c} CNR-VD's failure rate varies with noise levels and state preparation error rates. A white solid line at 0.5 failure rate marks CNR-VD's noise resilience limit, with a white dashed line showing the boundary for noisy VD. All the data in this graph presents average results from 50 experiments with random states, with error bars for standard deviation.}} 
  \label{fig3}
\end{figure}

\noindent\textbf{Numerical simulations}

To evaluate the performance of CNR-VD, we conduct various numerical simulations and compare its results among those obtained from noisy VD and other VD-based methods. To quantify the accuracy of estimators we use the absolute error calculated as 
$\text{Error}=|\langle O\rangle^{\text{est}}-\langle O\rangle^{\text{ideal}}|$.
For a fair comparison, we configure the same total measurement shot number for each method. We utilize second-order VD and perform the calibration procedure in each independent repeated experiment.

Our investigation begins with ascertaining the accuracy of CNR-VD. The stochastic Pauli gate noise model is utilized eliminating the need for RC technique. 
We define the noise level of the VD circuit as the magnitude of gate noise in the VD circuit, with the benchmark being the noise data from real machines~\cite{ZHU2022advantage} (i.e., when the noise level equals 1), and the value of the noise level indicates multiples of the benchmark noise value (See the detailed settings in the Supplementary Note 5). 

We generate random noisy state $\rho_r=(1-\epsilon)|\psi\rangle\langle\psi|+\epsilon|\psi_e\rangle\langle\psi_e|$ with the state preparation error rate $\epsilon$ by randomly choosing pure states $|\psi\rangle\langle\psi|$ and $|\psi_e\rangle\langle\psi_e|$ which are mutually orthogonal. This formulation yields a straightforward fidelity measure~\cite{Koczor2021Exponential, Cai2021quantumerror, Cai2021APractical} of $\mathcal{F}=\text{Tr}[\rho_r|\psi\rangle\langle\psi|]=1-\epsilon$, providing a clear and direct assessment of the noise impact on the fidelity of the prepared state relative to the ideal state $|\psi\rangle\langle\psi|$.

We first carry out simulations of 4-qubit random states with a fixed state preparation error rate $\epsilon=0.10$ and varying noise levels from 0.1 to 10.0, in comparison with ideal VD and noisy VD as illustrated in Fig.~\ref{fig3} \textbf{a}. Results demonstrate that CNR-VD well recovers the outcomes obtained by ideal VD for low-weight observables or moderate noise levels. Our subsequent focus lays in the noise threshold at which the error-mitigated methods provide enhanced results compared to unmitigated estimation. With reference to the diagram in Fig.~\ref{fig3} \textbf{b}, we indicate the corresponding noise thresholds with dashed lines. The increased noise threshold highlights the superiority of CNR-VD over noisy VD.
In addition, we evaluate the failure rate of CNR-VD contingent on fluctuating noise levels and state preparation error rates. The failure rate signifies the probability that the results generated by CNR-VD are inferior to unmitigated estimation. Since the accuracy of unmitigated estimation is merely determined by the state preparation error, larger state preparation error rate relatively to the noise level of the VD circuit tends to result in smaller failure rate, in accordance with the data shown in Fig.~\ref{fig3} \textbf{c}.
A boundary at failure rate 0.5 for CNR-VD is marked by a solid white line, signalling the point where CNR-VD’s resilience to noise diminishes. 
The corresponding failure rate boundary for the noisy VD is also highlighted by a white dashed line. From our observations, CNR-VD extends the boundary to higher noise level by up to one order of magnitude, which provides vital insights into the effectiveness and robustness of CNR-VD.

We also conduct a comparative study between CNR-VD and two common VD-based methods designed to counter noise in the VD circuit, which are Shadow Distillation~\cite{Alireza2022Shadow} (SD) and noisy VD combined with zero-noise extrapolation (ZNE-VD)~\cite{Koczor2021Exponential}. SD is a variation of VD that leverages shadow estimation to approximate the nonlinear function $\text{Tr}[\rho^2O]/\text{Tr}[\rho^2]$ of $\rho$. And ZNE-VD employs the zero-noise extrapolation (ZNE)~\cite{Endo2018Practical,Temme2017Error,Li2017Efficient,Cai2021multi,Hamilton2020Errormitigated,Lowe2021Unified,Garmon2020Benchmarking,Samuele2022Efficiently} QEM scheme to mitigate the noise in the VD circuit. The comparisons involve $N$-qubit random quantum states, which are generated by parameterized quantum circuits within the context of the Trotterized 1D Transverse-field Ising Model~\cite{Smith2019Simulating,Haoran2023Machine} (See the detailed settings in the Supplementary Note 5.). In our simulations, we fixed the observable to be single Z Pauli observable $\sigma_z^{N}$ on the last $N$-th qubit. For both CNR-VD and ZNE-VD, we employ RC technique with 4 random instances in the VD circuit to enhance the effectiveness of the estimators.
\begin{figure}[t]
  \centering
  {\includegraphics[width=0.48\textwidth,trim=0 0 0 0, clip]{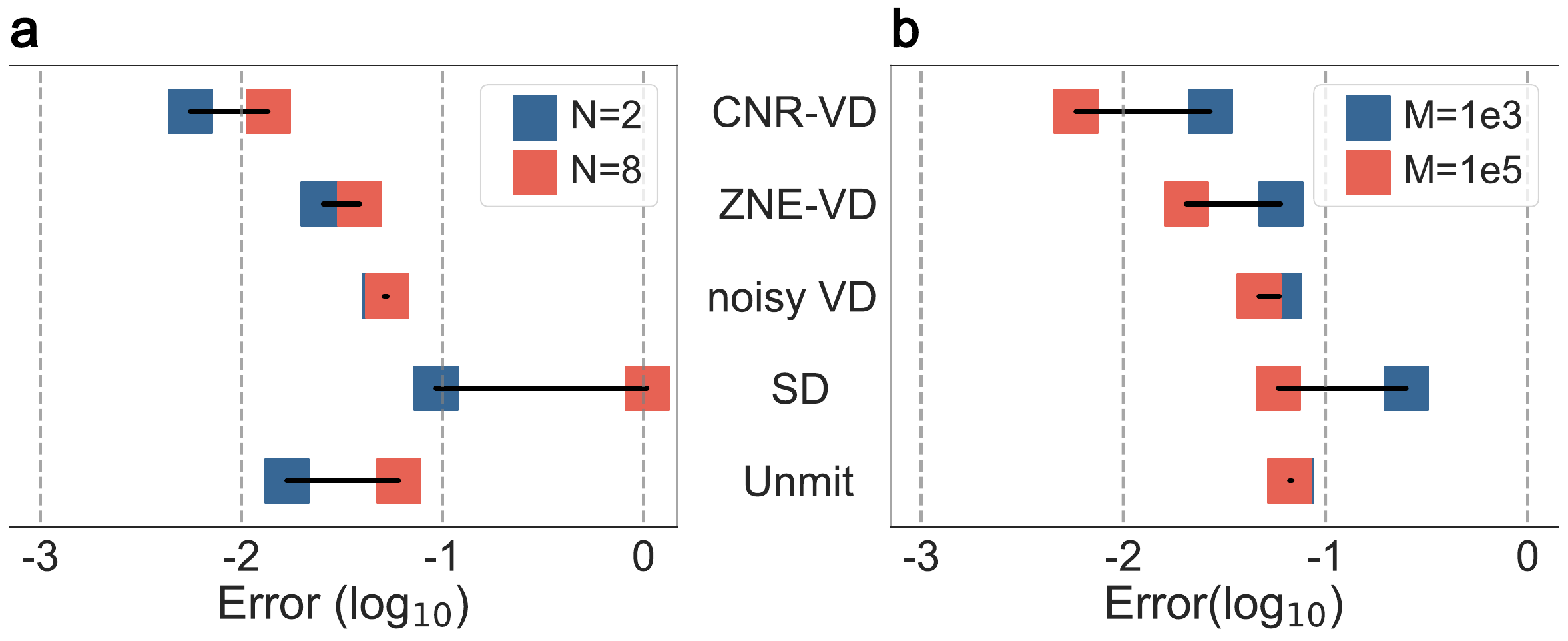}}
  \caption{{\textbf{Comparison results of the expectation values of random parameterized states.} \textbf{a} Performance of five estimators: CNR-VD, ZNE-VD,  noisy VD, SD and unmitigated estimation (Unmit) with fixed measurement shots number $M=2\times 10^4$, and number of qubits 2 and 8. \textbf{b} Performance of estimators for 4-qubit random parameterized state, with measurement shots number $10^3$ and $10^5$. The displayed data represents the average outcome derived from 20 separate experiments for each qubit number and circuit depth. And the chart includes error bars indicating the standard deviation.}}
  \label{fig4}
\end{figure}
We assess the accuracy across a range of qubit numbers $N$ from 2 to 8 and corresponding total measurement numbers $M$ spanning from $10^3$ to $10^5$. By specifically examining the accuracy at the extremes of these ranges, that is, using the maximum and minimum values of $N$ and $M$, we aim to provide a clear illustration of the trend, highlighting how the algorithm's performance scales with the system size and measurement resources.
Results are given in Fig.~\ref{fig4}.
We can observe that CNR-VD remains its superiority in accuracy even as the number of qubits increases. Notably, CNR-VD achieves an improvement in accuracy compared to noisy VD by more than an order of magnitude, as given in Fig.~\ref{fig4} \textbf{a}. And in Fig.~\ref{fig4} \textbf{b}, we can see that with the increasing number of shots, the performance of three VD-based methods (CNR-VD, ZNE-VD, SD) achieve significant enhancement in accuracy while the accuracy of noisy VD and unmitigated estimation are unable to further improve. It is noteworthy that among the three VD-based methods, CNR-VD consistently maintains a clear advantage.

Furthermore, the Supplementary Note 5 presents simulation outcomes for higher-order VD using CNR-VD, demonstrating its capacity to recover the ideal behavior of third-order VD under moderate noise conditions. Considering practical scenarios involving circuit compilation, our investigation is extended within the Supplementary Note 5 to assess the performance of VD-based algorithms, including the original VD, CNR-VD, and ZNE-VD, both pre- and post-compilation. The findings in the Supplementary Note 5 indicate that, despite some variation post-compilation, CNR-VD maintains its superiority in performance among the VD-based algorithms.

\noindent\textbf{\large{Discussion}}

In conclusion, we propose the CNR-VD which is established based on the original noisy VD and effectively overcomes the potential harmful influence of gate noise in the VD circuit. The simulation results demonstrate that our proposed method can achieve better mitigation performance under diverse noisy conditions, in comparison to noisy VD and other VD-based methods designed against noise in the VD circuit. Combined with randomized compiling technique we can also increase the threshold for noise level at which QEM algorithms outperform unmitigated estimations.

Moreover, since our theoretical analysis is not limited to the VD algorithm, the CNR-VD estimator can also provide an effective solution to the practical challenge of circuit noise for the Hadamard-Test algorithms. This opens up vast research opportunities for our future work, such as exploring advanced techniques of CNR-VD in more complex error environments, and investigating the practical implementation of our approach on real quantum computing platforms. We anticipate that this study may contribute valuable insights to the field of QEM and inform future research directions.

\noindent\textbf{\large{Methods}}

\noindent\textbf{The implementation of the randomized compiling employed in the CNR-VD}

\begin{table}
  \caption{The chosen subset of 3-qubit Pauli operators which are preserved by CSWAP gate under conjugation. Here the $q_i$ ($i=0,1,2$) denotes the $i-$th qubit of the CSWAP gate, and $I,X,Y,Z$ represent the identity, Pauli $\sigma_x$, Pauli $\sigma_y$ and Pauli $\sigma_z$ operators, respectively.}
  \resizebox{0.18\textwidth}{!}{
    \begin{tabular}{lll|lll}
    \hline  \hline
    $q_0$ & $q_1$ & $q_2$ & $q_0$ & $q_1$ & $q_2$ \\ \hline
    $I$  & $I$  & $I$  & $Z$  & $I$  & $I$  \\ \hline
    $I$  & $X$  & $X$  & $Z$  & $X$  & $X$  \\ \hline
    $I$  & $Y$  & $Y$  & $Z$  & $Y$  & $Y$  \\ \hline
    $I$  & $Z$  & $Z$  & $Z$  & $Z$  & $Z$  \\ \hline  \hline
    \end{tabular}
  }\label{tab-sub}
\end{table}
We employ randomized compilation (RC) technique in computing the CNR-VD estimator to come close to the main assumption of stochastic Pauli gate noise. In this section, we will provide a detailed explanation of the procedure.

Traditional RC is typically performed before and after the noisy multi-qubit gates, and newly generated circuits exhibit logical equivalence with their original counterparts. In practice, it is customary to select a fixed number of circuit with different random twirling instances and perform a certain number of measurements for each circuit, rather than resampling random twirling instance for per measurement shot.
\begin{figure}[t]
    \centering
    {\includegraphics[width=0.45\textwidth]{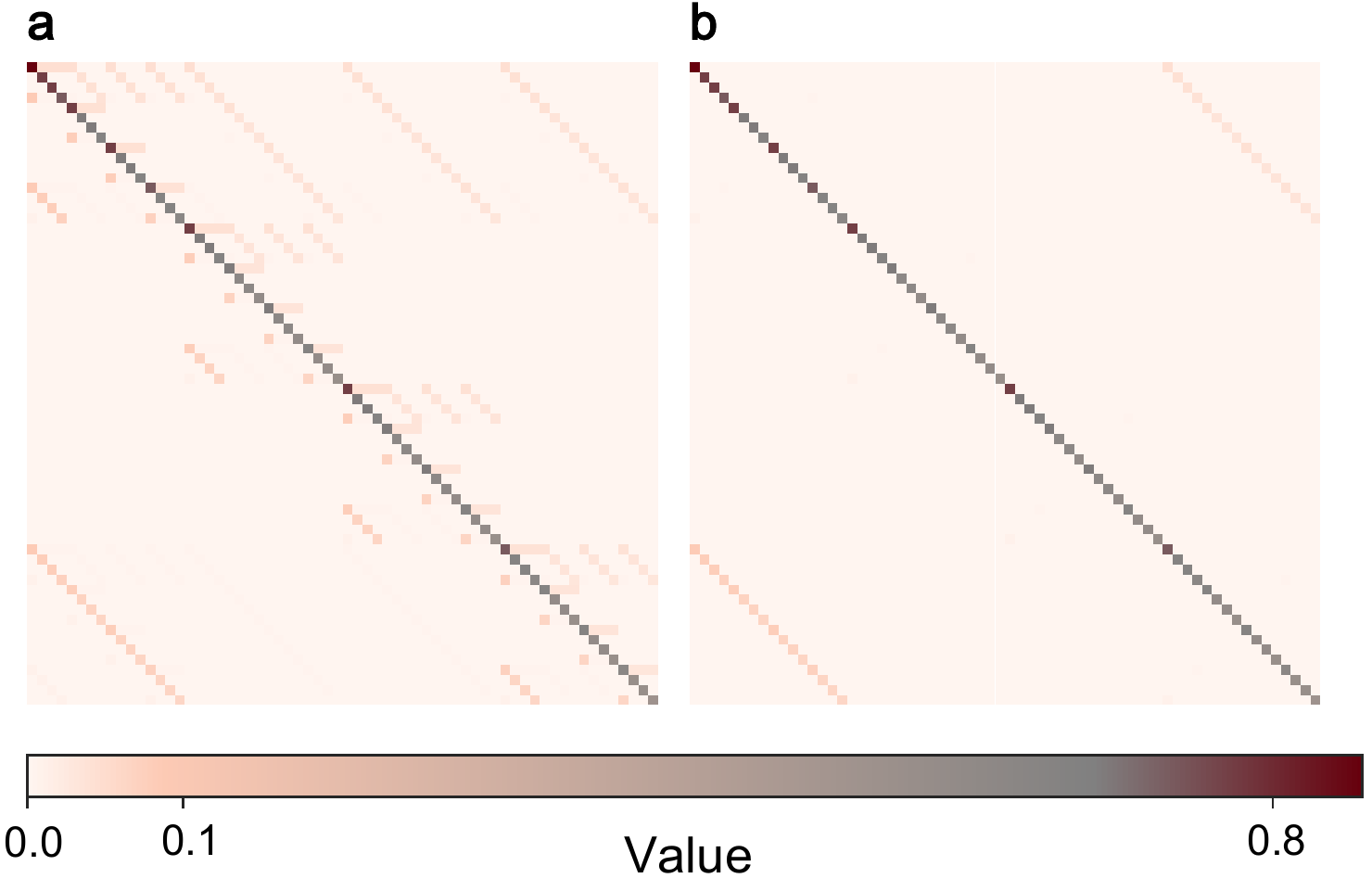}}
    \caption{{{\textbf{The effect of randomized compiling (RC) with the chosen subset of 3-qubit Pauli operators which are preserved by CSWAP gate under conjugation.} \textbf{a} The original Pauli Transfer Matrix (PTM)~\cite{Greenbaum2015aqm} $\mathcal{P}$ for the composite noise channel $\Lambda$ consisting of depolarizing noise and amplitude damping noise. The elements in the heat-map represent the magnitude of PTM, which are calculated using $\mathcal{P}_{i,j}=\frac{1}{8}\text{Tr}[P_i\Lambda(P_j)]$ where $P_i$ denotes the $i-$th 3-qubit Pauli operator. The off-diagonal elements of the PTM represent the non-Pauli components, which will be eliminated via RC. \textbf{b} The PTM for the composite noise channel after RC with the chosen subset of 3-qubit Pauli operators which are preserved by CSWAP gate under conjugation. It can be observed that the off-diagonal elements have been effectively suppressed, although only a small subset is selected.}}}\label{fig5}
\end{figure}
\begin{figure*}[t]
  \centering
  {\includegraphics[width=0.8\textwidth]{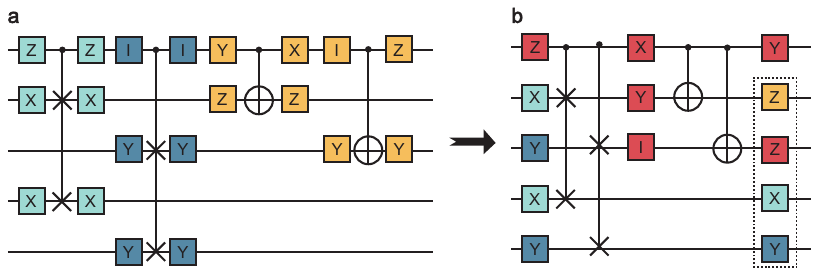}}
  \caption{{\textbf{The compilation of Pauli twirling gates for second-order VD circuit.
  } \textbf{a} The complete VD circuit with Pauli twirling gates. We have labeled the Pauli gates with different colors to indicate those associated with the same multi-qubit gate. \textbf{b} The VD circuit after compiling the Pauli twirling gates. The dashed rectangle in panel (b) signifies that the Pauli gates within it, corresponding to qubits other than the first, are not required to be executed, given that the VD algorithm only measures the first qubit. } }\label{fig6}
\end{figure*}

We will use second-order VD for all subsequent discussions in this section. We can perform RC on the CSWAP chain and controlled-$O$ chain separately. For the controlled-$O$ chain which is Clifford, we can randomly select a $2N+1$-qubit Pauli operator $P$ and calculate the corresponding conjugated Pauli gate $P^\text{twrl}$ by
\begin{equation}
    P^\text{twrl} = CO\cdot P\cdot CO^\dagger,
\end{equation}
where $CO$ represents the controlled-$O$ chain. When it comes to the CSWAP chain, as the CSWAP gate are not Clifford, the Pauli operator conjugated by the CSWAP gate may not still be a Pauli operator. We choose 8 Pauli gates from 3-qubit Pauli group in Tab.~\ref{tab-sub} which remain unchanged after being conjugated by CSWAP gate. Despite selecting only a small subset, RC yields promising results as shown in Fig.~\ref{fig5}.
Since the CSWAP gates in the CSWAP chain only share the control qubit, we can readily derive the form of Pauli operators preserved by the CSWAP gate under conjugation 
\begin{equation}
    T=T^\text{twrl}=\{I,Z\}^0\otimes\sum_{j=0}^{N-1}P^{j+1}P^{j+1+N},
\end{equation} 
where $P$ is randomly chosen Pauli operator and the superscript denotes the qubit on which the operator acts. 
It is possible to compile the Pauli twirling gates chosen for the CSWAP gate with the surrounding Pauli gates, due to the property of remaining unchanged under conjugation with the CSWAP gate. After compilation only three additional layers of Pauli gates are needed for implementing RC in the VD circuit, as illustrated in Fig.~\ref{fig6}. Additionally, the first layer of Pauli gates in the right figure can be compiled together with the single-qubit gates in the state preparation circuit of the input state. In the third layer of Pauli gates, it is only necessary to apply the Pauli gate on the first qubit, as we are only performing measurements on this qubit. 

\noindent\textbf{\large{Data availability}}

The data that support the plots within this paper and other findings of this study are available from the corresponding author upon reasonable request.

\noindent\textbf{\large{Code availability}}

The code used for producing the results is available on the Zenodo repository~\cite{zenodo}.

\noindent\textbf{\large{Author Contributions}}

H.-L. H. supervised the whole project. X.-Y. X and H.-L. H. conceived the idea and co-wrote the paper. X.-Y. X provided the theoretical analysis and proofs. X.-Y. X, C. D. and H.-L. H. carried out the numerical simulations and analyzed the results. All authors contributed to discussions of the results.

\noindent\textbf{\large{Acknowledgments}}

H.-L. H. acknowledges support from the National Natural Science Foundation of China (Grant No. 12274464), and Natural Science Foundation of Henan (Grant No. 242300421049). We gratefully acknowledge Hefei Advanced Computing Center for hardware support with the numerical experiments.

\bibliographystyle{naturemag}
\bibliography{b}

\end{document}


\title{Supplemental Information for ``Circuit-Noise-Resilient Virtual Distillation"}
\maketitle
\section{Virtual Distillation}
For an $N$-qubit noisy state $\rho$, the complete spectral decomposition of it can be expressed as 
\begin{equation}\label{eq-spectral}
    \rho=(1-\epsilon)|\psi\rangle\langle\psi|+\epsilon\sum_k\lambda|\psi_k\rangle\langle\psi_k|,
\end{equation}
where $0<\epsilon<1$ is the state preparation error rate and $|\psi\rangle\langle\psi|$ is the dominant eigenstate of $\rho$, along with the summation $\sum_k$ which satisfies $\sum_k\lambda_k=1$ taking over all the noisy components of $\rho$. 
Given $n$ identical copies of $\rho$, the error-mitigated expectation value of the Pauli observable $O$ via virtual distillation (VD) is
\begin{equation}
    \begin{aligned}
        \langle O\rangle_\text{VD}=\frac{\text{Tr}[\rho^nO]}{\text{Tr}[\rho^n]}=\frac{(1-\epsilon)^n\langle\psi|O|\psi\rangle+\epsilon^n\sum_k\lambda_k^n\langle\psi_k|O|\psi_k\rangle}{(1-\epsilon)^n+\epsilon^n\sum_k\lambda_k^n}=\langle\psi|O|\psi\rangle+\mathcal{O}(\epsilon^n),
    \end{aligned}
\end{equation}
which indicates that the approximation error $\mathcal{O}(\epsilon^n)$ decays exponentially with the number of copies $n$. To efficiently estimate Tr$[\rho^nO]$ (Tr[$\rho^n$]), we utilize the VD circuit which consists of the controlled derangment operator $D_n$ and the controlled operator $O$. 
The $n$-order VD circuit involves a three-qubit CSWAP gate chain $\prod_{t=0}^{n-1}\prod_{i=0}^{N-1}\text{CSWAP}_{i,t}$ where $\text{CSWAP}_{i,t}$ acts on qubit $Q_0$, $Q_{i+1+tN}$ and $Q_{i+1+(t+1)N}$, and a sequence of two-qubit controlled Pauli gates $\prod_{j=1}^{k}\text{CO}_{i_j}$ where $\text{CO}_{i_j}$ is applied on qubit $Q_0$ and $Q_{{i_j}+1}$. And we denote the set of index of qubits on the nontrivial Pauli terms of the $k$-weight Pauli observable $O$ as $\mathcal{I}_k$.   

We denote the probability of measuring the ancilla qubit $Q_0$ in the zero state as $p_0(\rho,O)$ and the $\text{Tr}[\rho^nO]$ can be estimated by $2p_0(\rho,O)-1$. Take the second-order VD ($n=2$) for example, due to the linear properties of quantum mechanics, we can rewrite the probability $p_0(\rho,O)$ according to the spectral decomposition
\begin{equation}
    p_0(\rho,O)=(1-\epsilon)^np_0(|\psi\rangle^{\otimes 2},O)+2\epsilon(1-\epsilon)\sum_k\lambda_kp_0(|\psi\rangle|\psi_k\rangle,O)+\epsilon^2\sum_{k,k'}\lambda_k\lambda_{k'}p_0(|\psi_k\rangle|\psi_{k'}\rangle,O).
\end{equation}
In this way, we can focus on probability of the pure state in form of $p_0(|\psi_a\rangle|\psi_b\rangle,O)$. The input state of the VD circuit is $|+\rangle|\psi_a\rangle|\psi_b\rangle$, and we can obtain the final ($2N+1$)-qubit density matrix $\rho^\text{out}$ just before measurement by
\begin{equation}
    \begin{aligned}
    \rho^\text{fin}=&\frac{1}{2}(
        |+\rangle|\psi_a\rangle|\psi_b\rangle\langle+|\langle\psi_a|\langle\psi_b| \ +\ |-\rangle O|\psi_b\rangle|\psi_a\rangle\langle-|\langle\psi_b|O\langle\psi_a|\ + \\
        &|+\rangle|\psi_a\rangle|\psi_b\rangle\langle-|\langle\psi_b|O\langle\psi_a| \ +\ |-\rangle O|\psi_b\rangle|\psi_a\rangle\langle+|\langle\psi_a|\langle\psi_b|).    
    \end{aligned}
\end{equation}
Then we calculate the probability $p_0(|\psi_a\rangle|\psi_b\rangle,O)$ via
\[
p_0(|\psi_a\rangle|\psi_b\rangle,O)=\text{Tr}[\rho^\text{fin}\cdot(|0\rangle\langle 0|\otimes I_{2N})]= \left\{
    \begin{array}{ll}
        \frac{1}{2},\  \text{if} \ a\neq b,\\ \\
        \frac{1}{2}+\frac{1}{2}\text{Tr}[|\psi_a\rangle\langle\psi_a|O], \ \text{if} \ a=b,
    \end{array}
  \right.
\]
where the derangement operator $D_2$ filters the asymmetric components of $\rho^{\otimes 2}$ to include only permutation-symmetric parts such as $|\psi_a\rangle|\psi_a\rangle$. Then we can arrive at
\begin{equation}
    \begin{aligned}
        2p_0(\rho,O)-1=&(1-\epsilon)^2(1+\langle\psi|O|\psi\rangle)+2\epsilon(1-\epsilon)\sum_k+\epsilon^2\sum_{k,k'}\lambda_k\lambda_{k'}+\epsilon^2\sum_k\langle|\psi_k|O|\psi_k\rangle-1\\
        =&(1-\epsilon)^2-1+\text{Tr}[\rho^2O]+2\sum_k\lambda_k\epsilon(1-\epsilon)+\epsilon^2\sum_{k,k'}\lambda_k\lambda_{k'}\\
        =&\text{Tr}[\rho^2O].
    \end{aligned}
\end{equation}
\begin{figure}[h]
    \centering
    {\includegraphics[width=0.95\textwidth]{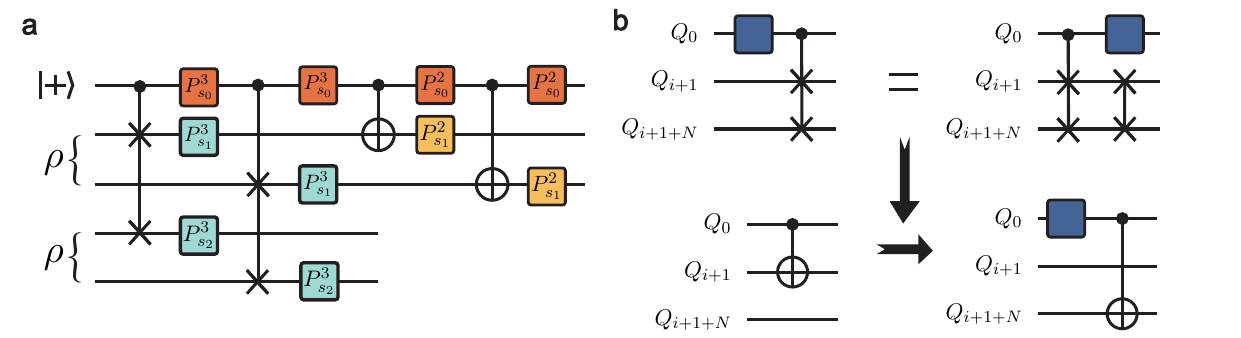}}
    \caption{{\textbf{The illustration of the pauli gate noise and its effect in the second-order VD circuit.}} \textbf{a} The 2-qubit and 3-qubit Pauli gate noise in the second-order VD circuit. \textbf{b} The effect of the $\sigma_x$ or $\sigma_y$ Pauli error (blue square) acting on the controlled qubit $Q_0$ of the CSWAP gate where we denote the $j$-th qubit of the VD circuit as $Q_j$. As $\left(\sigma_x(\sigma_y)\otimes I\otimes I\right)\cdot \text{CSWAP}=\text{CSWAP}\cdot\left(\sigma_x(\sigma_y)\otimes\text{SWAP}\right)$ which introduces a SWAP gate in the qubit $Q_{i+1}$ and $Q_{i+1+N}$, it leads to the transformation of controlled Pauli gate CO$_{i+1}$ to CO$_{i+1+N}$.}\label{fig-pauli}
\end{figure}

\section{Noisy Virtual Distillation under Pauli gate noise}
In the maintext, we consider the stochastic Pauli gate noise 
$
    \mathcal{E}^\text{P}_m(\rho)=(1-p_m)\rho+p_m\sum_{s=1}^{4^m-1}\delta_s^mP_s^m\rho P_s^m,
$
where $p_m\geq 0$ is the Pauli error rate and the coefficients $\{\delta_s^m\}_s$ of the $m$-qubit Pauli error terms $\{P_s^m\}_s$ satisfies $\delta_s^m\geq 0, \sum_s\delta_s^m=1$. We denote the $m$-qubit ($m=1,2,3$) Pauli error term as $P^m_s=\sum_{j=0}^{m-1}P^m_{s_j}$ where $P^m_{s_j}$ is from the single-qubit Pauli group $\{I,\sigma_x,\sigma_y,\sigma_z\}^{\otimes m}$. We now will provide the details of the noisy virtual distillation (VD) under Pauli gate noise in the VD circuit in this section. 

We assume the local Pauli gate noise model, which means that the noise introduced by a gate only affects the qubits on which the gate is applied. For simplificity, we assume that the single-qubit Pauli gate noise can be absorbed into the nearest multi-qubit gate noise. As a result, only 2-qubit Pauli gate noise affects controlled Pauli gates, and 3-qubit Pauli gate noise affects CSWAP gates, respectively. We define the noisy CSWAP gate $\widetilde{\text{CSWAP}}_{i,t}$ and controlled Pauli gate $\widetilde{\text{CO}}_{i_j}$ as
\begin{equation}
    \widetilde{\text{CSWAP}}_{i,t}=\mathcal{E}^\text{P}_3\cdot\text{CSWAP}_{i,t}, \ \ \widetilde{\text{CO}}_{i_j}=\mathcal{E}^\text{P}_2\cdot\text{CO}_{i_j},
\end{equation}
as illustrated in the Fig.~\ref{fig-pauli} a). As we solely measure the ancilla qubit $Q_0$, it is essential to consider how the Pauli gate noise in the VD circuit affects the measurement outcomes of $Q_0$. 

\text{\it{The noise in the CSWAP gate chain}}--- As the $N$ CSWAP gates in the VD circuit only share their controlled qubit $Q_0$, only the first qubit of the Pauli error term $P^3_{s_0}$ can influence its following CSWAP gates. Then we can investigate the impact of Pauli terms in the form of $P^3_{s_0}\otimes I\otimes I$ before the CSWAP gate,
\[
\text{CSWAP}\cdot\left(P^3_{s_0}\otimes I\otimes I\right)= \left\{
    \begin{array}{ll}
        \left(P^3_{s_0}\otimes I\otimes I\right)\cdot\text{CSWAP},\  \text{if} \ P^3_{s_0}=\sigma_z,I\\ \\
        \left(P^3_{s_0}\otimes \text{SWAP}\right)\cdot\text{CSWAP}, \ \text{if} \ P^3_{s_0}=\sigma_x,\sigma_y,
    \end{array}
  \right.
\]
which means that the $\sigma_x$ or $\sigma_y$ Pauli error occurring on the ancilla qubit $Q_0$ can generate a undesired SWAP gate on the data qubits for each CSWAP gate behind it. Due to the light-cone principle, not all erroneous SWAP gates will affect the measurement outcome. When the SWAP gate acting on $(i+1,i+1+N)$ meets $i\in \mathcal{I}_k$, it will lead to the transformation of the corresponding controlled Pauli gate from CO$_{i+1}$ to CO$_{i+1+N}$, as demonstrated in Fig.~\ref{fig-pauli} b). This can further result in a portion of single-qubit Pauli terms of $O$, which originally act only on the register corresponding to the first copy of $\rho$, also affecting the second register. We define $O=O_1+O_2$ and $i=1,2$ refers to the index of the register which the operator $O_i$ acts on. Then the final density matirx turns to
\begin{equation}
    \begin{aligned}
        \tilde{\rho}^\text{fin}=&\frac{1}{2}(|+\rangle|\psi_a\rangle|\psi_b\rangle\langle+|\langle\psi_a|\langle\psi_b| \ + \ |-\rangle O_1|\psi_b\rangle O_2|\psi_a\rangle\langle -|\langle\psi_b|O_1\langle\psi_a|O_2 \ + \\
        &|+\rangle|\psi_a\rangle|\psi_b\rangle\langle -|\langle\psi_b|O_1\langle\psi_a|O_2 \ + \ |-\rangle O_1|\psi_b\rangle O_2|\psi_a\rangle\langle +|\langle\psi_a|\langle\psi_b|),
    \end{aligned}
\end{equation}
and the noisy probability $\tilde{p}_0(|\psi_a\rangle|\psi_a\rangle,O)$ becomes
\[
\tilde{p}_0(|\psi_a\rangle|\psi_b\rangle,O)=\text{Tr}[\tilde{\rho}^\text{fin}\cdot(|0\rangle\langle 0|\otimes I_{2N})]= \left\{
    \begin{array}{ll}
        \frac{1}{2},\  \text{if} \ a\neq b,\\ \\
        \frac{1}{2}+\frac{1}{2}\text{Tr}[|\psi_a\rangle\langle\psi_a|O_1]\cdot\text{Tr}[|\psi_a\rangle\langle\psi_a|O_2], \ \text{if} \ a=b.
    \end{array}
\right.
\]\label{eq-pauli1}
Hence, the noisy estimation of Tr$[\rho^2O]$ contaminated by the undesired SWAP gate caused by the $\sigma_x$ or $\sigma_y$ Pauli error on the ancilla is
\begin{equation}
    \begin{aligned}
        2\tilde{p}_0(\rho,O)-1=&(1-\epsilon)^2(1+\text{Tr}[|\psi\rangle\langle\psi|O_1]\cdot\text{Tr}[|\psi\rangle\langle\psi|O_2])+2\epsilon(1-\epsilon)\sum_k\lambda_k+\\
        &\epsilon^2\sum_{k,k'}\lambda_k\lambda_{k'}+\epsilon^2\sum_k\lambda_k\text{Tr}[|\psi_k\rangle\langle\psi_k|O_1]\cdot\text{Tr}[|\psi_k\rangle\langle\psi_k|O_2]-1\\
        =&(1-\epsilon)^2\text{Tr}[|\psi\rangle\langle\psi|O_1]\cdot\text{Tr}[|\psi\rangle\langle\psi|O_2]+\epsilon^2\sum_k\lambda_k\text{Tr}[|\psi_k\rangle\langle\psi_k|O_1]\cdot\text{Tr}[|\psi_k\rangle\langle\psi_k|O_2].
    \end{aligned}
\end{equation}

\text{\it{The noise in the controlled Pauli gate chain}}--- We then turn our attention to the pauli gate noise in the controlled Pauli gates, along with the residual Pauli noise in the CSWAP gate chain which commutes with the CSWAP gate chain. As the controlled Pauli gate chain is Clifford which conserve the Pauli group, we can serve these Pauli errors as a ($2N+1$)-qubit stochastic Pauli noise channel which acts after the final density matrix $\rho^\text{fin}$. We consider the effect of the single-qubit Pauli error term $P_{s_0}$, it gives
\[
\tilde{p}_0(|\psi_a\rangle|\psi_b\rangle,O)= \left\{
    \begin{array}{ll}
        \frac{1}{2},\  \text{if} \ a\neq b,\ \text{or} \ a=b\ \text{and} \ P_{s_0}=\sigma_y,\\ \\
        \frac{1}{2}+\frac{1}{2}\text{Tr}[|\psi_a\rangle\langle\psi_a|O], \ \text{if} \ a=b\ \text{and} \ P_{s_0}=I,\sigma_x, \\  \\
        \frac{1}{2}+\frac{1}{2}\text{Tr}[|\psi_a\rangle\langle\psi_a|O], \ \text{if} \ a=b\ \text{and} \ P_{s_0}=\sigma_z,
    \end{array}
  \right.
\]
and then the noisy estimation is 
\[
2\tilde{p}_0(\rho,O)-1= \left\{
    \begin{array}{ll}
        \text{Tr}[\rho^2O],\  \text{if} \ P_{s_0}=I,\sigma_x,\\ \\
        0, \ \text{if} \ P_{s_0}=\sigma_y, \\  \\
        -\text{Tr}[\rho^2O], \ \text{if}\ P_{s_0}=\sigma_z.
    \end{array}
\right. \equiv a\text{Tr}[\rho^2O],
\]
where the parameter $a$ is $\rho$-independent and only determined by the structure of the VD circuit and the noise level of the Pauli gate noise in the VD circuit.

\text{\it{The outcome of noisy virtual distillation}}--- From above discussions, it can be inferred that the undesired SWAP errors due to the $\sigma_x,\sigma_y$ errors on the ancilla qubit $Q_0$ can lead to a partial shift of the Pauli observable $O$ from register 1 to register 2, while the remaining Pauli errors result in multiplying the shifted expectation value by a parameter independent of $\rho$. When the erroneously generated SWAP gates separates $O$ into $O_1$ and $O_2$, which reside in registers 1 and 2 respectively, we can obtain a similar expression for $\tilde{p}_0(|\psi_a\rangle|\psi_a\rangle,O)$ as
\[
\tilde{p}_0(|\psi_a\rangle|\psi_a\rangle,O)= \left\{
    \begin{array}{ll}
        \frac{1}{2}+\frac{1}{2}\langle\psi_a|O_1|\psi_a\rangle\cdot\langle\psi_a|O_2|\psi_a\rangle,\ \text{if} \ P_{s_0}=I,\sigma_x,\\ \\
        \frac{1}{2},\ \text{if} \ P_{s_0}=\sigma_y, \\  \\
        \frac{1}{2}-\frac{1}{2}\langle\psi_a|O_1|\psi_a\rangle\cdot\langle\psi_a|O_2|\psi_a\rangle, \ \text{if} \ P_{s_0}=\sigma_z, 
    \end{array}
  \right.
\]
However, the $\sigma_x$ and $\sigma_y$ errors on the ancilla qubit $Q_0$ also cancel each other out, making it difficult to infer the number of the unwanted SWAPs generated based on the occurrences of $\sigma_x,\sigma_y$ errors. Therefore, we combine similar terms of expectation values resulting from different Pauli error terms based on the number of unwanted SWAPs in the VD circuit. 
Then the outcome obtained from the noisy VD circuit can be represented as a function that
\begin{equation}\label{eq-noise2}
    2\tilde{p}_0(\rho,O)-1=\sum_{j=0}^{\lceil\frac{k}{2}\rceil}\sum_{t=0}^{\binom{k}{j}-1}a^k_{j,t}(p_1,p_2,p_3,O)A_j(t),
\end{equation}
where $j$ represents the total number of SWAP errors that lead to nontrivial terms of $O$ being transferred out of register 1, corresponding to $\binom{k}{j}$ situations. And $t$ refers to the indices of these situations. It is worth noting that when $k$ is even and $j=k/2$, the number of all the possible combinations of $O_1$ and $O_2$ is $\binom{k}{j}/2$. We do not make specific mention of this here in order to maintain the uniformity in the expression of $2\tilde{p}_0(\rho,O)-1$. 

\section{Circuit-Noise-Resilient virtual distillation under general Markovian noise model}
\subsection{The proof for the full resilience to noise in the VD circuit for the CNR-VD}
In the maintext we have defined the effective noise $\mathcal{E}^\text{eff}$ of the VD circuit, then the noisy probability $\tilde{p}_0(\rho,O)$ of measuring zero state in the ancilla is
\begin{equation}\label{eqs1}
    \tilde{p}_0(\rho,O)=\text{Tr}[\mathcal{E}^\text{eff}\mathcal{U}(|+\rangle\langle+|\otimes\rho^{\otimes n})(|0\rangle\langle 0|\otimes I_{nN})],
\end{equation}
here the operator $\mathcal{U}$ refers to the controlled-$D_n$, controlled-$O$ and the Hadamard gate $H$. We have defined the CNR-VD estimator as
\begin{equation}\label{eq-est2}
  \hat{O}_\text{CNR-VD}(\rho)=\frac{\hat{O}_\text{VD}(\rho)}{\hat{O}_\text{VD}(s)},
\end{equation}
when the effective noise $\mathcal{E}^\text{eff}$ satisfies any of these two conditions:
\begin{itemize}
  \item There exists a set of Kraus operators such that: each Kraus operator $K_i=K_i^\text{R}\otimes K_i^\text{D}$ of $\mathcal{E}^\text{eff}$ is separate into subsystems $\rho^\text{R}$ and $\rho^\text{D}$, along with unitary $K_i^\text{R}$ and $K_i^\text{D}$.
  \item $\mathcal{E}^\text{eff}=\mathcal{E}^\text{eff,R}\otimes\mathcal{E}^\text{eff,D}$, along with the single-qubit unital noise channel $\mathcal{E}^\text{eff,R}$ and the CPTP noise channel $\mathcal{E}^\text{eff,D}$,
\end{itemize}
then the CNR-VD estimator $\hat{O}_\text{CNR-VD}(\rho)$ can reduce the error in the expectation value indued by $\mathcal{E}^\text{eff}$.

{\text{\it{Proof for Condition 1.}}}---For simplificity, we suppose the state preparation error for $s$ is negligible compared to $\mathcal{E}^\text{eff}$. And we take second-order VD and we assume that this two copies of noisy state are the same. Then the tensor product of the state $\rho^{\otimes 2}$ can be written as a linear combination of four types of pure state terms as shown in Eq.~\ref{eq-spectral}: $|\psi\rangle|\psi\rangle$, $|\psi\rangle|\psi_k\rangle$, $|\psi_k\rangle|\psi\rangle$ and $|\psi_k\rangle|\psi_{k'}\rangle$. Then we can decompose the $\tilde{p}_0(\rho,O)$ in Eq.~\ref{eqs1} as $\tilde{p}_0(\rho,O)=(1-\epsilon)^2\tilde{p}_0(|\psi\rangle|\psi\rangle,O)+\epsilon(1-\epsilon)\sum_k\lambda_k[\tilde{p}_0(|\psi\rangle|\psi_k\rangle,O)+\tilde{p}_0(|\psi_k\rangle|\psi\rangle,O)]+\epsilon^2\sum_{k,k'}\lambda_k\lambda_{k'}\tilde{p}_0(|\psi_k\rangle|\psi_{k'}\rangle,O)$. 
When the effective noise $\mathcal{E}^\text{eff}$ is unital noise channel and each Kraus operator conforms to $K_i=K_i^\text{R}\otimes K_i^\text{D}$ along with unitary $K_i^\text{R}$ and $K_i^\text{D}$, then the expression of $\tilde{p}_0(|\psi\rangle\langle\psi|^{\otimes 2},O)$ in Eq.~\ref{eq-pp} turns to
\begin{equation}
    \begin{aligned}
        \tilde{p}_0(|\psi\rangle|\psi\rangle,O)&=\frac{1}{2}\text{Tr}[\mathcal{E}^{\text{eff}}\left(\mathcal{U}(|+\rangle|\psi\rangle|\psi\rangle)\cdot \mathcal{U}(|+\rangle|\psi\rangle|\psi\rangle)^\dagger\right)\cdot(|0\rangle\langle 0|\otimes I_{2N})]\\
        &=\frac{1}{2}\text{Tr}[\mathcal{E}^\text{eff}(|+\rangle\langle+|\otimes |\psi\rangle|\psi\rangle\langle\psi|\langle\psi|)\cdot(|0\rangle\langle 0|\otimes I_{2N})]+\frac{1}{2}\text{Tr}[\mathcal{E}^\text{eff}(|-\rangle\langle-|\otimes |\psi\rangle O|\psi\rangle\langle\psi|O\langle\psi|) \\
        &\cdot(|0\rangle\langle 0|\otimes I_{2N})]+\frac{1}{2}\text{Tr}[\mathcal{E}^\text{eff}(|+\rangle\langle-|\otimes |\psi\rangle|\psi\rangle\langle\psi|O\langle\psi|)\cdot(|0\rangle\langle 0|\otimes I_{2N})]\\
        &+\frac{1}{2}\text{Tr}[\mathcal{E}^\text{eff}(|-\rangle\langle+|\otimes |\psi\rangle O|\psi\rangle\langle\psi|\langle\psi|)\cdot(|0\rangle\langle 0|\otimes I_{2N})]\\
        &=\frac{1}{2}\sum_i\text{Tr}[(K_i^\text{R}|+\rangle\langle +|{K_i^\text{R}}^\dagger)|0\rangle\langle 0|]\cdot\text{Tr}[K_i^\text{D}|\psi\rangle\langle\psi|^{\otimes 2}{K_i^\text{D}}^\dagger]\\
        &+\frac{1}{2}\sum_i\text{Tr}[(K_i^\text{R}|-\rangle\langle -|{K_i^\text{R}}^\dagger)|0\rangle\langle 0|]\cdot\text{Tr}[K_i^\text{D}(O|\psi\rangle\langle\psi|O)^{\otimes 2}{K_i^\text{D}}^\dagger]\\
        &+\frac{1}{2}\sum_i\text{Tr}[(K_i^\text{R}|+\rangle\langle -|{K_i^\text{R}}^\dagger)|0\rangle\langle 0|]\cdot\text{Tr}[K_i^\text{D}((|\psi\rangle\langle\psi|O)\otimes|\psi\rangle\langle\psi|){K_i^\text{D}}^\dagger]\\
        &+\frac{1}{2}\sum_i\text{Tr}[(K_i^\text{R}|-\rangle\langle +|{K_i^\text{R}}^\dagger)|0\rangle\langle 0|]\cdot\text{Tr}[K_i^\text{D}(|\psi\rangle\langle\psi|\otimes(|\psi\rangle\langle\psi|O)){K_i^\text{D}}^\dagger]\\
        &=\frac{1}{2}\sum_i\text{Tr}[(K_i^\text{R}I{K_i^\text{R}}^\dagger)|0\rangle\langle 0|]+\frac{1}{2}\text{Tr}[|\psi\rangle\langle\psi|O]\sum_i\text{Tr}[(K_i^\text{R}Z{K_i^\text{R}}^\dagger)|0\rangle\langle 0|]\\
        &:=\frac{1}{2}+b\text{Tr}[|\psi\rangle\langle\psi|O],
    \end{aligned}
\end{equation}
where we use the fact that the unitary Kraus operator meeting $K_i^\text{R}{K_i^\text{R}}^\dagger=I$. And for other three pure state terms, the same approach can be employed for computation, yielding the following result:
\begin{equation}
    \begin{aligned}
        &\tilde{p}_0(|\psi\rangle|\psi_k\rangle,O)=a,\\
        &\tilde{p}_0(|\psi_k\rangle|\psi\rangle,O)=a,\\
    \end{aligned}
\end{equation}
\[
\tilde{p}_0(|\psi_k\rangle|\psi_{k'}\rangle,O)= \left\{
    \begin{array}{ll}
        a,  \text{if} \ k\neq k'\\ \\
        a+b\text{Tr}[|\psi_k\rangle\langle\psi_k|O], \ \text{else}.
    \end{array}
  \right.
\]
Thus we can obtain the expression of $\tilde{p}_0(\rho,O)$ as 
\begin{equation}
    \begin{aligned}\label{eq-CNR1}
      \tilde{p}_0(\rho,O)&=(1-\epsilon)^2(\frac{1}{2}+b\text{Tr}[|\psi\rangle\langle\psi|O])+\epsilon(1-\epsilon)+\epsilon^2(\frac{1}{2}+b\sum_k\lambda_k^2\text{Tr}[|\psi_k\rangle\langle\psi_k|O])  \\
      &=\frac{1}{2}+b\left((1-\epsilon)^2\text{Tr}[|\psi\rangle\langle\psi|O]+\epsilon^2\sum_k\lambda_k^2\text{Tr}[|\psi_k\rangle\langle\psi_k|O]\right)\\
      &=\frac{1}{2}+b\text{Tr}[\rho^2O].
    \end{aligned}
\end{equation} 
Then we can deduce that 
\begin{equation}
    \frac{2\tilde{p}_0(\rho,O)-1}{2\tilde{p}_0(s,O)-1}=\text{Tr}[\rho^2O].
\end{equation}
Similarly, it can be obtained that 
\begin{equation}
    \frac{2\tilde{p}_0(\rho,I)-1}{2\tilde{p}_0(s,I)-1}=\text{Tr}[\rho^2].
\end{equation}
Then in this situation, the simplified CNR-VD estimator comply with
\begin{equation}\label{eq-p2}
    \hat{O}_\text{CNR-VD}=\frac{2\tilde{p}_0(\rho,O)-1}{2\tilde{p}_0(\rho,I)-1}\cdot\frac{2\tilde{p}_0(s,O)-1}{2\tilde{p}_0(s,I)-1}=\frac{\hat{O}_\text{VD}(s)}{\hat{O}_\text{VD}(\rho)}=\frac{\text{Tr}[\rho^2O]}{\text{Tr}[\rho^2]},
\end{equation}
which can also be easily extended to $n>2$.

{\text{\it{Proof for Condition 2.}}}---When the effective noise $\mathcal{E}^\text{eff}$ is separate on two subsystems R and D that $\mathcal{E}^\text{eff}=\mathcal{E}^\text{eff,R}\otimes \mathcal{E}^\text{eff,D}$ along with the single-qubit unital noise channel $\mathcal{E}^\text{eff,R}$ and CPTP noise channel $\mathcal{E}^\text{eff,D}$ , then this yields
\begin{equation}
    \begin{aligned}
        \tilde{p}_0(|\psi\rangle|\psi\rangle,O)&=\frac{1}{2}\text{Tr}[\mathcal{E}^\text{eff,R}(|+\rangle\langle +|)|0\rangle\langle 0|]\cdot\text{Tr}[\mathcal{E}^\text{eff,D}(|\psi\rangle\langle\psi|^{\otimes 2})] \\
        &+\frac{1}{2}\text{Tr}[\mathcal{E}^\text{eff,R}(|-\rangle\langle -|)|0\rangle\langle 0|]\cdot\text{Tr}[\mathcal{E}^\text{eff,D}((O|\psi\rangle\langle\psi|O)^{\otimes 2})]\\
        &+\frac{1}{2}\text{Tr}[\mathcal{E}^\text{eff,R}(|+\rangle\langle -|)|0\rangle\langle 0|]\cdot\text{Tr}[\mathcal{E}^\text{eff,D}((|\psi\rangle\langle\psi|O)\otimes|\psi\rangle\langle\psi|)] \\
        &+\frac{1}{2}\text{Tr}[\mathcal{E}^\text{eff,R}(|-\rangle\langle +|)|0\rangle\langle 0|]\cdot\text{Tr}[\mathcal{E}^\text{eff,D}(|\psi\rangle\langle\psi|\otimes(|\psi\rangle\langle\psi|O))] \\
        &=\frac{1}{2}\text{Tr}[\mathcal{E}^\text{eff,R}(I)|0\rangle\langle 0|]+\frac{1}{2}\text{Tr}[\mathcal{E}^\text{eff,D}(Z)|0\rangle\langle 0|]\cdot\text{Tr}[|\psi\rangle\langle\psi|O] \\
        &=\frac{1}{2}+b\text{Tr}[|\psi\rangle\langle\psi|O],
    \end{aligned}
\end{equation}
where we use the fact $\text{Tr}[\mathcal{E}^\text{eff,D}(\sigma)]=\text{Tr}[\sigma]$ for the CPTP noise channel $\mathcal{E}^\text{eff,D}$ and any operator $\sigma$, and $\mathcal{E}^\text{eff,R}(I)=I$ for the unital noise channel $\mathcal{E}^\text{eff,R}$. Therefore, through a similar deduction process as condition 1, we can conclude that the Eq.~\ref{eq-p2} also holds under condition 2.

\subsection{CNR-VD estimator with loosen requirement of the effective noise}
We also establish other form of the CNR-VD estimator with relaxed requirement. For differentiation, we denote the $s^+$ and $s^-$ as the eigenstate of Pauli $O$ with eigenvalue $+1$ and $-1$, respectively.
As for $O=I$ where only eigenvalue $+1$ exists, we replace the first copy of $s^+$ with $g^0$ ($g^\frac{1}{2}$) which satisfies $|\langle s^+|g^0\rangle|^2=0$ (and $|\langle s^+|g^\frac{1}{2}\rangle|^2=1/2$), and the noisy probability is referred to $\tilde{p}_0(g^0,I)$ ($\tilde{p}_0(g^\frac{1}{2},I)$). 
We will utilize these noisy probabilities obtained from both noisy VD and the calibration stage to establish the CNR-VD estimator in the form of 
\begin{equation}\label{eq-sm-est1}
  \begin{aligned}
      \hat{O}_{\text{CNR-VD}}(\rho)=\left(\frac{2\tilde{p}_0(\rho,O)-\tilde{p}_0(s^+,O)-\tilde{p}_0(s^-,O)}{\tilde{p}_0(\rho,I)-\tilde{p}_0(g^0,I)}\right)
      \times\left(\frac{2(\tilde{p}_0(g^\frac{1}{2},I)-\tilde{p}_0(g^0,I))}{\tilde{p}_0(s^+,O)-\tilde{p}_0(s^-,O)}\right).
    \end{aligned}
\end{equation}

We now turn our attention to the precise condition when $\hat{O}_{\text{CNR-VD}}(\rho)$ exhibits full resilience to the effective noise. We can get the following condition. 
\begin{theorem}
If the effective noise $\mathcal{E}^\text{eff}$ satisfys
\begin{equation}\label{eq-noise1}
  \begin{aligned}
    \text{Tr}_{1:nN}[\mathcal{E}^\text{eff}(\rho^\text{RD})]=\sum_ic_i\text{Tr}[\rho^\text{D}]\rho^\text{R}_i, 
    \ \  \text{with} \  \rho^\text{R}_i=\text{Tr}_{1:nN}[K_i(\rho^R\otimes I_{nN})K_i^\dagger],
  \end{aligned}
\end{equation}
where $\text{Tr}_{1:nN}$ denotes the partial trace over qubits from index 1 to $nN$ and $c_i$ is some constant which is exclusively associated with $\mathcal{E}^\text{eff}$, then the CNR-VD estimator $\hat{O}_{\text{CNR-VD}}(\rho)$ given in Eq.~\ref{eq-sm-est1}
can eliminate the undesired bias in the expectation value caused by $\mathcal{E}^\text{eff}$.
\end{theorem}
The criteria of $\mathcal{E}^\text{eff}$ set out in Eq.~\ref{eq-noise1} means that the effective noise does not introduce entanglement between $\rho^\text{R}$ and $\rho^\text{D}$.

\text{\it{Proof.}}--- 
For the effective noise $\mathcal{E}^\text{eff}$ satisfying the conditions described in theorem 1, $\tilde{p}_0(|\psi\rangle|\psi\rangle,O)$ equals
\begin{equation}\label{eq-pp}
    \begin{aligned}
        \tilde{p}_0(|\psi\rangle|\psi\rangle,O)&=\frac{1}{2}\text{Tr}[\mathcal{E}^{\text{eff}}\left(\mathcal{U}(|+\rangle|\psi\rangle|\psi\rangle)\cdot \mathcal{U}(|+\rangle|\psi\rangle|\psi\rangle)^\dagger\right)\cdot(|0\rangle\langle 0|\otimes I_{2N})]\\
        &=\frac{1}{2}\text{Tr}[\mathcal{E}^\text{eff}(|+\rangle\langle+|\otimes |\psi\rangle|\psi\rangle\langle\psi|\langle\psi|)\cdot(|0\rangle\langle 0|\otimes I_{2N})]+\frac{1}{2}\text{Tr}[\mathcal{E}^\text{eff}(|-\rangle\langle-|\otimes |\psi\rangle O|\psi\rangle\langle\psi|O\langle\psi|) \\
        &\cdot(|0\rangle\langle 0|\otimes I_{2N})]+\frac{1}{2}\text{Tr}[\mathcal{E}^\text{eff}(|+\rangle\langle-|\otimes |\psi\rangle|\psi\rangle\langle\psi|O\langle\psi|)\cdot(|0\rangle\langle 0|\otimes I_{2N})]\\
        &+\frac{1}{2}\text{Tr}[\mathcal{E}^\text{eff}(|-\rangle\langle+|\otimes |\psi\rangle O|\psi\rangle\langle\psi|\langle\psi|)\cdot(|0\rangle\langle 0|\otimes I_{2N})]\\
        &=\frac{1}{2}\sum_ic_i\text{Tr}[K_i((|+\rangle\langle+|+|-\rangle\langle-|)\otimes I_{2N})K_i^\dagger\cdot (|0\rangle\langle 0|\otimes I_{2N})]\\
        &+\frac{1}{2}\text{Tr}[|\psi\rangle\langle\psi|O]\cdot\sum_ic_i\text{Tr}[K_i((|+\rangle\langle-|+|-\rangle\langle+|)\otimes I_{2N})K_i^\dagger\cdot (|0\rangle\langle 0|\otimes I_{2N})]\\
        &=\frac{1}{2}\sum_ic_i\text{Tr}[K_iK_i^\dagger\cdot(|0\rangle\langle 0|\otimes I_{2N})]+\frac{1}{2}\text{Tr}[|\psi\rangle\langle\psi|O]\cdot\sum_ic_i\text{Tr}[K_i((Z\otimes I_{2N})K_i^\dagger\cdot (|0\rangle\langle 0|\otimes I_{2N})] \\
        &:=a+b\text{Tr}[|\psi\rangle\langle\psi|O],
    \end{aligned}
\end{equation}
where the noise parameters $a$ and $b$ are unrelated to the input state $|\psi\rangle\langle\psi|^{\otimes 2}$ since $c_i$ is only exclusively relative with $\mathcal{E}^\text{eff}$.
Thus we can obtain the expression of $\tilde{p}_0(\rho,O)$ in the similar way as in Eq.~\ref{eq-CNR1} 
\begin{equation}
    \begin{aligned}
      \tilde{p}_0(\rho,O)&=(1-\epsilon)^2(a+b\text{Tr}[|\psi\rangle\langle\psi|O])+2a\epsilon(1-\epsilon)+\epsilon^2(a+b\sum_k\lambda_k^2\text{Tr}[|\psi_\rangle\langle\psi_k|O])  \\
      &=a+b\left((1-\epsilon)^2\text{Tr}[|\psi\rangle\langle\psi|O]+\epsilon^2\sum_k\lambda_k^2\text{Tr}[|\psi_\rangle\langle\psi_k|O]\right)\\
      &=a+b\text{Tr}[\rho^2O].
    \end{aligned}
\end{equation} 
When we input the eigenstate $s^+$ ($s^-$) of the Pauli observable $O$ with eigenvalue 1 (-1), we can obtain $\tilde{p}_0(s^+,O)=a+b$ ($\tilde{p}_0(s^-,O)=a-b$). As a result, it follows that
\begin{equation}
    \frac{2\tilde{p}_0(\rho,O)-\tilde{p}_0(s^+,O)-\tilde{p}_0(s^-,O)}{\tilde{p}_0(s^+,O)-\tilde{p}_0(s^-,O)}=\text{Tr}[\rho^2O].
\end{equation} 
For $O=I$, the $\tilde{p}_0(\rho,I)=a'+b'\text{Tr}[\rho^2]$ and $\tilde{p}_0(s^+,I)=a'+b'$. When we replace the first copy of $s^+$ with $g^0$, since $s^+$ and $g^0$ are both pure states we can get 
\begin{equation}
    \tilde{p}_0(g^0,I)=a'+b'|\langle s^+|g^0\rangle|^2=a',
\end{equation}
following Eq.~\ref{eq-pp}. And the same way can be used to obatin $\tilde{p}_0(g^\frac{1}{2},I)=a'+b'/2.$ Then we get
\begin{equation}
    \frac{\tilde{p}_0(\rho,I)-\tilde{p}_0(g^0,I)}{2(\tilde{p}_0(g^\frac{1}{2},I)-\tilde{p}_0(g^0,I))}=\text{Tr}[\rho^2],
\end{equation}
which means the CNR-VD estimator defined in Theorem 1 satisfies
\begin{equation}
    \begin{aligned}
        \hat{O}_\text{CNR-VD}(\rho)&=\frac{2\tilde{p}_0(\rho,O)-\tilde{p}_0(s^+,O)-\tilde{p}_0(s^-,O)}{\tilde{p}_0(\rho,I)-\tilde{p}_0(g^0,I)}\\
        &\cdot \frac{2(\tilde{p}_0(g^\frac{1}{2},I)-\tilde{p}_0(g^0,I))}{\tilde{p}_0(s^+,O)-\tilde{p}_0(s^-,O)}=\frac{\text{Tr}[\rho^2O]}{\text{Tr}[\rho^2]}.
    \end{aligned}
\end{equation}
And this can be easily extended to $n>2$ by expanding $\rho^n$ in terms of the binomial series.

\subsection{The variance of the CNR-VD estimator}
Since the CNR-VD estimator is in a form of quotient, we can employ the error propagation formula to calculate the variance of the CNR-VD estimator, incorporating the first-order Taylor expansion into the calculation. Consider a function $Y = A/B$, where $A$ and $B$ are random variables. When $A$ and $B$ are mutual independent, we can compute the variance of $Y$ denoted as Var$[Y]$ by
\begin{equation}
  \text{Var}[Y]\approx \left(\frac{\mathbb{E}[A]}{\mathbb{E}[B]}\right)^2\cdot \left(\frac{\text{Var}[A]}{(\mathbb{E}[A])^2}+\frac{\text{Var}[B]}{(\mathbb{E}[B])^2} \right).
\end{equation}

Here we only calculate the variance of the CNR-VD estimator in Eq.~\ref{eq-est2} and we can compute the variance of other version of CNR-VD estimator in the similar way. The variance of the simpified $\hat{O}_\text{CNR-VD}(\rho)$ is
\begin{equation}
  \begin{aligned}
    &\text{Var}[\hat{O}_\text{CNR-VD}(\rho)]=\text{Var}[\frac{\hat{O}_\text{VD}(\rho)}{\hat{O}_\text{VD}(s)}]\\
    &=\left(\frac{\mathbb{E}[\hat{O}_\text{VD}(\rho)]}{\mathbb{E}[\hat{O}_\text{VD}(s)}\right)^2 \left(\frac{\text{Var}[\hat{O}_\text{VD}(\rho)]}{(\mathbb{E}[\hat{O}_\text{VD}(\rho)])^2}+\frac{\text{Var}[\hat{O}_\text{VD}(s)]}{(\mathbb{E}[\hat{O}_\text{VD}(s)])^2} \right).
  \end{aligned}
\end{equation}
If we assume $\hat{O}_\text{VD}(\rho)$ has the same variance for any input state under the same measurement shots number $M$, we can infer that the variance of $\hat{O}_\text{CNR-VD}(\rho)$ scales proportionally to that of $\hat{O}_\text{VD}(\rho)$ by
\begin{equation}
  \begin{aligned}
    &\text{Var}[\hat{O}_\text{CNR-VD}(\rho)]=\\
    &\left(\frac{1}{(\mathbb{E}[\hat{O}_\text{VD}(s)])^2}+\frac{(\mathbb{E}[\hat{O}_\text{VD}(\rho)])^2}{(\mathbb{E}[\hat{O}_\text{VD}(s)])^4}\right)\text{Var}[\hat{O}_\text{VD}(\rho)]\\
    &\leq \left(\frac{(\mathbb{E}[\hat{O}_\text{VD}(s)])^2+1}{(\mathbb{E}[\hat{O}_\text{VD}(s)])^4}\right)\text{Var}[\hat{O}_\text{VD}(\rho)],
  \end{aligned}
\end{equation}
where the factor is totally determined by $\mathbb{E}[\hat{O}_\text{VD}(s)]$ which indicates the noise level in the VD circuit. 

To investigate the relationship among $\mathbb{E}[\hat{O}_\text{VD}(s)]$, the qubit number $N$, and the order $n$ more intuitively, we consider the simplest scenario where only local depolarizing gate noise with error rate $\gamma$ occurs in the circuit. Since the depolarizing noise meets $\mathcal{E}^\text{depo}_\gamma(P)=(1-\frac{4}{3}\gamma)P$ for non-trivial Pauli terms X, Y and Z, we can get $\mathbb{E}[\hat{O}_\text{VD}(s)]=\frac{(1-\frac{4}{3}\gamma)^{(n-1)N+k}}{(1-\frac{4}{3}\gamma)^{(n-1)N}}=(1-\frac{4}{3}\gamma)^{k}$ and $k$ is weight of observable $O$ which is the count of non-trivial Pauli terms in $O$. Consequently, we deduce
\begin{equation}
    \frac{(\mathbb{E}[\hat{O}_\text{VD}(s)])^2+1}{(\mathbb{E}[\hat{O}_\text{VD}(s)])^4}=\frac{(1-\frac{4}{3}\gamma)^{2k}+1}{(1-\frac{4}{3}\gamma)^{4k}}=\frac{1}{(1-\frac{4}{3}\gamma)^{2k}}+\frac{1}{(1-\frac{4}{3}\gamma)^{4k}}.
\end{equation}
We perform a Taylor expansion around $\gamma=0$, truncating the series at $o(\gamma^2)$,
\begin{equation}
    (1-\frac{4}{3}\gamma)^{-k}=1+\frac{4}{3}k\gamma+\frac{8k(k+1)}{9}\gamma^2.
\end{equation}
Thus, we arrive at
\begin{equation}
    \frac{(\mathbb{E}[\hat{O}_\text{VD}(s)])^2+1}{(\mathbb{E}[\hat{O}_\text{VD}(s)])^4}\approx 2+(8\gamma+\frac{16}{3}\gamma^2)k+\frac{160}{9}\gamma^2k^2,
\end{equation}
which entails that the scaling factor of variance for the CNR-VD estimator remains uncorrelated with the order of VD, and when the level of gate noise is minimal, the scaling factor exhibits a polynomial growth with the weight $k$ of the observable.

\section{Extension for general applications}
\begin{figure}[t]
    {\includegraphics[width=0.6\textwidth]{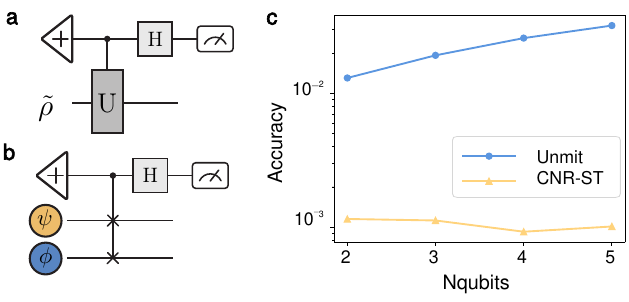}}
    \caption{{\textbf{Coalesced presentation showcasing the circuit of Hadamard-Test and SWAP-Test and the simulation results pertaining to the SWAP-Test.}} \textbf{a} The quantum circuit of Hadamard-Test. \textbf{b} The illustration of SWAP-Test to estimate the overlap of two quantum states $\psi$ and $\phi$. \textbf{c} The simulation results of SWAP-Test with unmitigated estimation (Unmit) and CNR-ST. The displayed data represents the average outcome derived from 50 separate experiments, with the error bars indicating the corresponding standard deviation.}\label{figs-st}
  \end{figure}

In order to demonstrate the generality of CNR-VD, we perform numerical simulations on a representative application of the Hadamard-Test, namely the SWAP-Test~\cite{Fanizza2020Beyond,Escartin2013SWAP}. SWAP-Test is a fundamental protocol in quantum computing. By swapping the two states via performing CSWAP operators, SWAP-Test enables to estimate the state overlap which provides valuable insights into quantum state comparison and discrimination. We conducted numerical experiments utilizing the SWAP-Test on two randomly generated quantum states $\psi$ and $\phi$ as illustrated in Fig.~\ref{figs-st} \textbf{b}. We call the SWAP-Test using the CNR estimator as CNR-ST, and the input state $s$ is $|0\rangle^{\otimes N}$ The accuary are still calculated using the absolute error.

To alleviate the impact of the extreme values of the ideal overlap on the estimation of absolute error, we generated the pair of random quantum states with a fixed overlap
\begin{equation}
  |\psi\rangle =U_\text{R}|0\rangle^{\otimes N},|\phi\rangle = U_\text{R}|+\rangle^{\otimes N},
\end{equation}
where the random unitary operator $U_\text{R}$ can maintain the inner product of $\psi$ and $\phi$ as $\langle\psi|\phi\rangle=\langle 0|^{\otimes N}U_\text{R}^\dagger U_\text{R}|+\rangle^{\otimes N}=\langle 0|^{\otimes N}|+\rangle^{\otimes N}=\frac{1}{2}$. 
We conduct simulations on estimating the overlap of random states on qubits with number ranging from 2 to 5 and compare the accuracy between the unmitigated SWAP-Test (Unmit) and CNR-ST. The results are given in Fig.~\ref{figs-st} \textbf{c}. This two approaches are implemented with the same number of total meaurement shots $10^5$. There is clearly an improvement in the accuracy of the estimated overlap with an order of magnitude, and the accuracy does not decrease significantly with an increasing number of qubits as unmitigated SWAP-Test does.

\section{{Details of numerical simulations}}
\subsection{The implementation of shadow distillation and VD integrated with zero-noise extrapolation}
\begin{figure}[t]
    \centering
    {\includegraphics[width=0.8\textwidth]{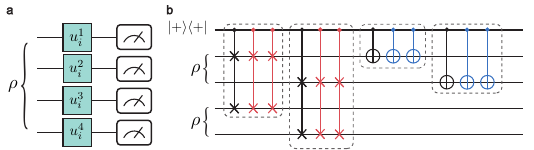}}
    \caption{{\textbf{The visualization presented depicts the process of 4-qubit shadow distillation (SD) and 2-qubit VD integrated with zero-noise extrapolation (ZNE-VD).
    } \textbf{a} Random single-qubit Clifford gate $u_i^{(j)}(j=1,2,3,4)$ is applied to each qubit of $\rho$, followed by the computation of $\frac{\text{Tr}[\rho^2X^{\otimes 2}]}{\text{Tr}[\rho^2]}$ through shadow estimation utilizing the measurement outcomes of a set of random gates $\{U_i=\prod_{j=1}^4u_i^{(j)}\}_i$. \textbf{b} The noise scaling method for zero-noise extrapolation is implemented by gate unfolding. Each multi-qubit gate within the VD circuit is executed three times in the noise-amplified circuit, while the dashed part in (b) is logically equivalent to its corresponding gate.}}
    \label{fig-twomethod}
\end{figure}

In the numerical simulations conducted in this maintext (shown in Fig.~4), we compare the CNR-VD with two common VD-based methods which are designed against noise in the VD circuit, namely Shadow Distillation and VD integrated with Zero-Noise Extrapolation. In this section, we will provide a comprehensive description of the implementation process for these two methods, and we only use second-order VD in this simulations. The illustration of these two methods are given in Fig.~\ref{fig-twomethod}.
\subsubsection{Shadow Distillation}
Shadow Distillation (SD) is one of the variants of VD which use shadow estimation technique to estimate the unlinear function $\frac{\text{Tr}[\rho^2O]}{\text{Tr}[\rho^2]}=\frac{\text{Tr}[D_2(\rho\otimes(\rho O))]}{\text{Tr}[D_2\rho^{\otimes 2}]}$ of $\rho$. In the SD, we treat the operator $D_2(I\otimes O)$ ($D_2$) as an observable. By randomly choosing $N_u$ local Clifford gates $\{U_i\}_i$, we obtain the corresponding shadow $\bar{\rho}_i$ from $N_s$ measuremeant outcomes for each $U_i$. Then we can use the estimator $\hat{O}_\text{SD}(\rho)=\hat{O}^\text{num}_\text{SD}(\rho)/\hat{O}^\text{den}_\text{SD}(\rho)$ to compute the error-mitiagted expectation value where  
\begin{equation}
    \hat{O}^\text{num}_\text{SD}(\rho)=\frac{1}{N_u(N_u-1)}\sum_{i\neq i'}\text{Tr}[D_2\bar{\rho}_i\otimes(\bar{\rho}_{i'}O)],
\end{equation}
and we can get $\hat{O}^\text{den}_\text{SD}(\rho)$ by replacing $O$ with $I$. In our simulations, we set $N_s=10$ if the total measurement shots $M\leq 10^4$, otherwise $N_s=50$. {And the number of snapshots $N_u$ is determined by the measurement shots number $M$, which is calculated by $N_u=\frac{M}{N_s}$.}

\subsubsection{VD integrated with Zero-Noise Extrapolation}
Typical quantum error mitigation methods such as Zero-Noise Extrapolation (ZNE) can also be employed to minimize the noise in the VD circuit, and we denote this method as ZNE-VD. ZNE employs data gathered across varying error rates to develop a model of expectation values, which can be extrapolated to zero noise limit. We use the Richardson extrapolation~\cite{Temme2017Error,Endo2018Practical} to calculate $\text{Tr}[\rho^2O]$ by
\begin{equation}
    \hat{O}_\text{ZNE-VD}^\text{num}(\rho)=\sum_{k=0}^m\gamma_k\hat{O}_\text{VD}^\text{num}(\rho,\lambda_k\epsilon),
\end{equation}
here we employ the denotation that $\hat{O}_\text{VD}^\text{num}(\rho,\lambda_k\epsilon)$ is the estimation of $\text{Tr}[\rho^2O]$ under error rate $\lambda_k\epsilon$ where $\lambda_k>=1$ is the amplified coefficient for the original noise level $\epsilon$. And the fitting coefficients $\{\gamma_k\}_k$ meet $\sum_{k=0}^m\gamma_k=1$ and $\sum_{k=0}^m\gamma_k\lambda_k^j=0$ for $j=1,...,m$ and the solutions gives $\gamma_k=\prod_{i\neq j}\frac{\lambda_j}{\lambda_j-\lambda_i}$. We can define the denumerator $\hat{O}_\text{ZNE-VD}^\text{dun}(\rho)$ estimating $\text{Tr}[\rho^2]$ in a similar way. Then the error-mitiagted expectation value of ZNE-VD is computed via $\hat{O}_\text{ZNE-VD}(\rho)=\hat{O}_\text{ZNE-VD}^\text{num}(\rho)/\hat{O}_\text{ZNE-VD}^\text{den}(\rho)$.

We choose the gate unfolding as the noise scaling methods for ZNE. Each multi-qubit gate in the VD circuit is applied $2m+1$ (for some integer $m$) times in the noise-amplified circuit, where $2m+1$ is the amplified coefficient and $m=1$ is chosen in our simulations. In the numerical simulations, we also employ RC technique in the ZNE to enhance the efficacy of gate unfolding applied on the VD circuit.

\subsection{{Circuit implementation and noise model in the numerical simulations}}

{{
    In the maintext, we conduct simulations under two scenarios. In the first scenario, we examine the characteristics of CNR-VD under the presence of random Pauli noise In the second scenario, we consider a more realistic situation, where we apply a unified noise model across both state preparation and the VD circuits.
}}


\subsubsection{{{Circuits for quantum simulations of 1D transverse-field Ising model}}}
\begin{figure}[t]
    {\includegraphics[width=0.32\textwidth]{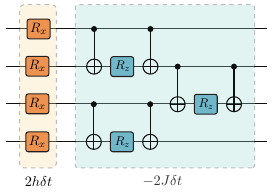}}
    \caption{{\textbf{The trotterized circuit for the quantum simulations of 1D transverse-field Ising model.} We set $\delta t=1$, then the angle of rotation-$x$ gate is $2h$ and the angle of rotation-$z$ gate is $-2J$.}}\label{figs-TS}
  \end{figure}
{The 1D transverse-field Ising model serves as a fundamental paradigm in quantum statistical mechanics, capturing essential aspects of phase transitions and critical phenomena. The temporal evolution within the spin chain is governed by the Hamiltonian,
\begin{equation}\label{eq-hising}
    \begin{aligned}
        H_\text{ising}&=-J\sum_{j=0}^{N-2}\sigma_z^j\sigma_z^{j+1}+h\sum_{j=0}^{N-1}\sigma_x^j\\
        &=H^Z_\text{ising}+H^X_\text{ising},
    \end{aligned}
\end{equation}
where we define $H^Z_\text{ising}=-J\sum_j\sigma_z^j\sigma_z^{j+1}$ and $H^X_\text{ising}=\sum_j\sigma_x^j$. $J$ and $h$ are the parameters which denote the coupling strength between neighboring spins and the transverse magnetic field, respectively. The qubits are initialized in zero state $|0\rangle^{\otimes N}$. }

{In our simulations, we employ the first-order Trotterized decomposition to approximate the time-evolution operator,
\begin{equation}
    \begin{aligned}
        e^{-iH_\text{ising}t}&=e^{-i(H^Z_\text{ising}+H^X_\text{ising})t}\\
        &\approx e^{-iH^Z_\text{ising}t}e^{-iH^X_\text{ising}t}.
    \end{aligned}
\end{equation}
For one time step $\delta t$, we have 
\begin{equation}
    \begin{aligned}
        &e^{-iH^Z_\text{ising}\delta t}=\prod_je^{-i(-J\delta t)\sigma_z^j\sigma_z^{j+1}}\\
        &e^{-iH^X_\text{ising}\delta t}=\prod_ie^{-i(h\delta t)\sigma_x^j},
    \end{aligned}
\end{equation}
and the circuit structure of a layer is shown in Fig.~\ref{figs-TS}. In the maintext, we randomly choose the parameters $J$ and the ratio $J/h$ from the interval $[0.05,0.2]$ and $[0.2,1.5]$ respectively. The step varies from 2 to 6 for each number of qubits ranging from 2 to 8.}

\subsubsection{Gate noise model}
We employ two types of gate noise model which are the stochastic Pauli gate noise model and the composite gate noise model consisting of depolarizing and amplitude damping.

\text{\it{The stochastic Pauli gate noise.}}---We set $\epsilon$ to be the noise level which is the average error rate for 2-qubit gates. The benchmark of the noise (when $\epsilon=1$) is from $Zuchongzhi$ 2.1~\cite{ZHU2022advantage}. With the benchmark $c_1=1.6\times 10^{-3}$ for single-qubit error and $c_2=6\times 10^{-3}$ for 2-qubit error, we can calculate the Pauli error rates $p_1=(\epsilon\cdot c_1/(1-1/2))\cdot(1-1/4)$ and $p_2=(\epsilon\cdot c_2/(1-1/4))\cdot(1-1/16)$~\cite{Arute2019supremacy} with randomly chosen Pauli noise parameters $\{\delta_s^2\}_s$. We use the same stochastic Pauli noise within each independent experiment including noisy VD and the calibration stage. Taking into consideration that the experimental implementation of the CSWAP gate typically involves the use of 2-qubit gates for compilation, we have adopted the compilation method outlined in Ref~\cite{Koczor2021Exponential}, where a minimum of six 2-qubit gates is required to compile the CSWAP gate. Consequently, according to Ref~\cite{Arute2019supremacy}, the process fidelity of the CSWAP gate is six times greater than that of 2-qubit gates under Pauli error rate $p_2$. As a result, we are able to calculate that the fidelity of the CSWAP gate is $\mathcal{F}(\text{CSWAP})=(1-p_2)^6$
and derive the corresponding Pauli error rate $p_3=1-\mathcal{F}(\text{CSWAP})=1-(1-p_2)^6$.

\text{\it{The composite gate noise.}}---Given the noise level $\epsilon$, we employ single-qubit depolarizing gate noise

\begin{equation*}
    \mathcal{E}_\text{depo}(\cdot)=(1-\epsilon\cdot p_\text{depo})(\cdot)+\frac{1}{2}\epsilon\cdot p_\text{depo}I_2,
\end{equation*}

for each qubit of the gate and the $p_\text{depo}$ is computed in the similar way as the Pauli gate noise~\cite{Arute2019supremacy}.
We also ultilize single-qubit amplitude damping $\mathcal{E}_\text{ampl}$ with the following definition

\begin{equation*}
    \mathcal{E}_\text{ampl}(\cdot)=E_0(\cdot)E_0^\dagger+E_1(\cdot)E_1^\dagger
\end{equation*}

where the Kraus operator $E_0=\begin{bmatrix}
    1 & 0 \\ 0 & \sqrt{1-\epsilon\cdot p_\text{ampl}} 
\end{bmatrix}$, $E_1=\begin{bmatrix}
    0 & \sqrt{\epsilon\cdot p_\text{ampl}} \\  0 & 0
\end{bmatrix}$. Here the $p_\text{ampl}$ is calculated using $p_\text{ampl}=1-e^{-\frac{d_2}{d_1}}$ where $d_1=22.67\times 10^{-6}$ and $d_2=20\times 10^{-9}$ represent the average data for the T1 and T2 duration collected from $Zuchongzhi$ 2.1~\cite{ZHU2022advantage,ding2023noise}. For the noise associated with multi-qubit gates, we employ the form of tensor product of single-qubit noise operators. In our simulations, we set the noise level $\epsilon=1$.
\begin{figure}[t]
    {\includegraphics[width=0.42\textwidth]{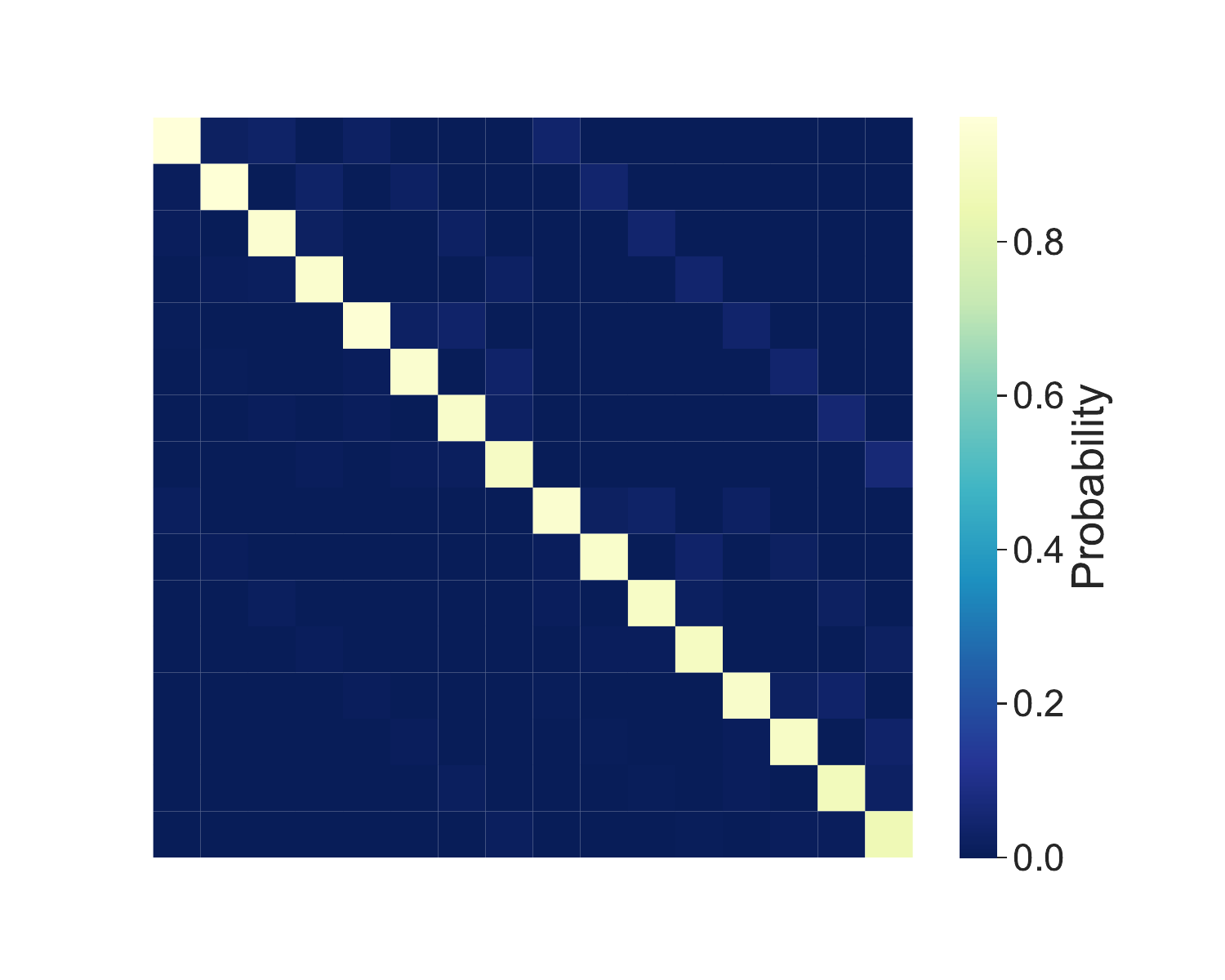}}
    \caption{{\textbf{The 4-qubit transfer matrix used in the measurement process.} The measurement data is obtained from $Zuchongzhi$ 2.1.}}\label{figs-transfer}
  \end{figure}

\subsubsection{Readout noise model}
It is common to assume the classical readout error model where the transfer matrix $\Lambda$ is employed to discribe the transformation between ideal and noisy measuremeant probability distributions which are denoted as $\mathbf{p}^\text{ideal}$ and $\mathbf{p}^\text{noisy}$
\begin{equation}\label{eq-readout}
    \mathbf{p}^\text{ideal}=\Lambda^{-1}\mathbf{p}^\text{noisy}.
\end{equation}
The element $\Lambda_{\mathbf{x},\mathbf{y}}$ of the response matrix $\Lambda$ is expressed as
\begin{equation}
    \Lambda_{\mathbf{x},\mathbf{y}}=\langle\mathbf{x}|\rho|\mathbf{y}\rangle, \ \ \mathbf{x},\mathbf{y}\in \mathbb{Z}_2^N,
\end{equation} 
where $\mathbf{x}\in \mathbb{Z}_2^N$ refers to the measuremeant outcome.
We utilize the Iterative Bayesian Unfolding (IBU) method~\cite{Nachman2020unfolding} to calculate $\mathbf{p}^\text{ideal}$ in order to avoid the direct inversion in Eq.~\ref{eq-readout}, which may produce unphysical probabilites and amplify the statistical uncertainties in $\Lambda$. We use 50 IBU iterations and iteration formula for the $(t+1)$-th iteration is calculated as follows
\begin{equation}
    m_i^{t+1}=\sum_j\frac{\Lambda_{ji}m_i^t}{\sum_k\Lambda_{jk}m_k^t}\times r_j,
\end{equation}
where $m_i$ and $r_j$ are the $i$-th and $j$-th terms in the probability estimated by IBU and the $\mathbf{p}^\text{noisy}$, respectively.

The full transfer matrix $\Lambda$ we used as readout error is built on the calibration data in $Zuchongzhi$ 2.1~\cite{ZHU2022advantage} as shown in Fig.~\ref{figs-transfer} and the estimated transfer matrix $\bar{\Lambda}$ for IBU is obtained under the Tensor Product Noise~\cite{Bravyi2021Mitigate} (TPN) model that $\bar{\Lambda}=\bar{\Lambda}_1\otimes\bar{\Lambda}_2\otimes...\otimes\bar{\Lambda}_N$. The sigle-qubit transfer matrix $\bar{\Lambda}_i$ on $(i-1)$-th qubit also relies on the calibration data in $Zuchongzhi$ 2.1. We use the TPN model because full matrix calibration is exponentially costly. However, this assumption completely ignores the correlated readout error, which leads to additional correlated measurement noise in SD and unmitigated estimation.

\begin{figure}[t]
    {\includegraphics[width=0.65\textwidth]{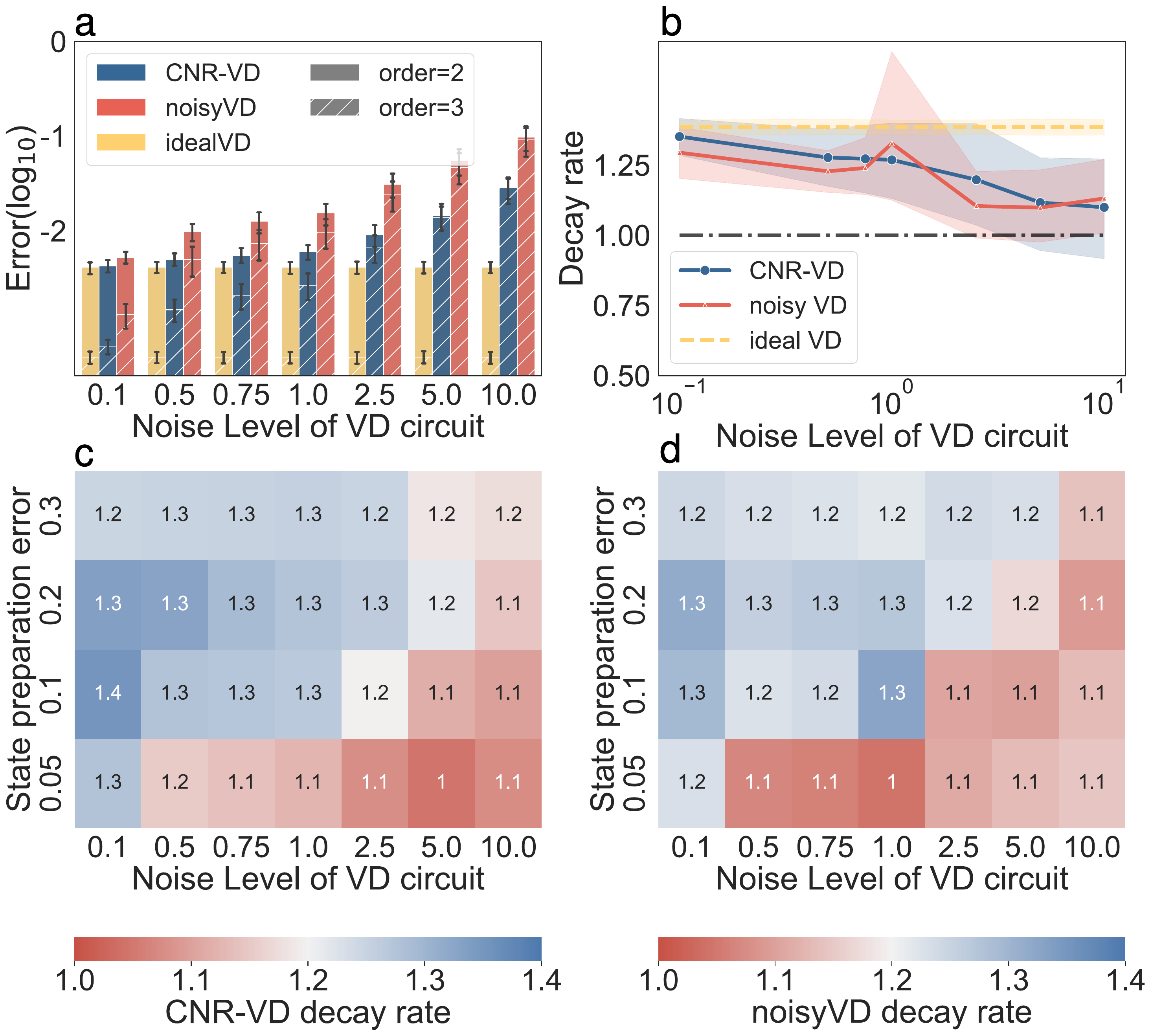}}
    \caption{{\textbf{Performance on CNR-VD of random noisy states, with different orders of VD.}} \textbf{a} The complete visualization of the variation in accuracy with order 2,3 and varying noise levels. \textbf{b} The decay rate of the expectation value estimation error varying with the noise levels. \textbf{c} The heat map of the decay rate for CNR-VD varying with state preparation error rates and noise levels of the VD circuit. \textbf{d} The heat map of the decay rate for noisy VD.
    The displayed data represents the average outcome derived from 50 separate experiments with different random states.}\label{figs-order}
  \end{figure}

\subsection{Relative simulation results}
\subsubsection{Simulation results for higher order VD}
We focus on the second-order VD in the maintext. Here, we supplement several simulations for 3-qubit random noisy state $\rho_r$ concerning higher-order VD and explore the relationship between the accuracy of CNR-VD and the increase in the order of VD with order 2 and order 3, noise level in the VD circuit changing from 0.1 to 10.0 and the state preparation error rate varying from $0.05$ to $0.30$. Referring to the data illustrated in Fig.~\ref{figs-order} \textbf{a}, CNR-VD demonstrates an impressive ability to recover the ideal behavior of VD, with noise level lower than 1. However, as the noise level increases, the incremental improvement in accuracy for CNR-VD from increasing the order diminishes gradually, reaching a point at 5 where the results for order 3 are actually inferior to those for order 2. We can also observe this from Fig.~\ref{figs-order} \textbf{b}, where we define the decay rate as the logarithm of the ratio of the accuary of order 2 to that of order 3. 
It is obvious that the cost-effectiveness of increasing the order becomes diminished as the noise level increases. We further present a detailed demonstration of the decay rate of CNR-VD and noisy VD as they vary with the changes in noise levels of the VD circuit and state preparation error rates, as shown in Fig.~\ref{figs-order} \textbf{c} and Fig.~\ref{figs-order} \textbf{d}. This implies that the improvement in error mitigation will not yield a continuous exponential benefit with increasing order.

\subsubsection{Detailed simulation results for random parameterized state}
\begin{figure}[t]
    {\includegraphics[width=0.65\textwidth]{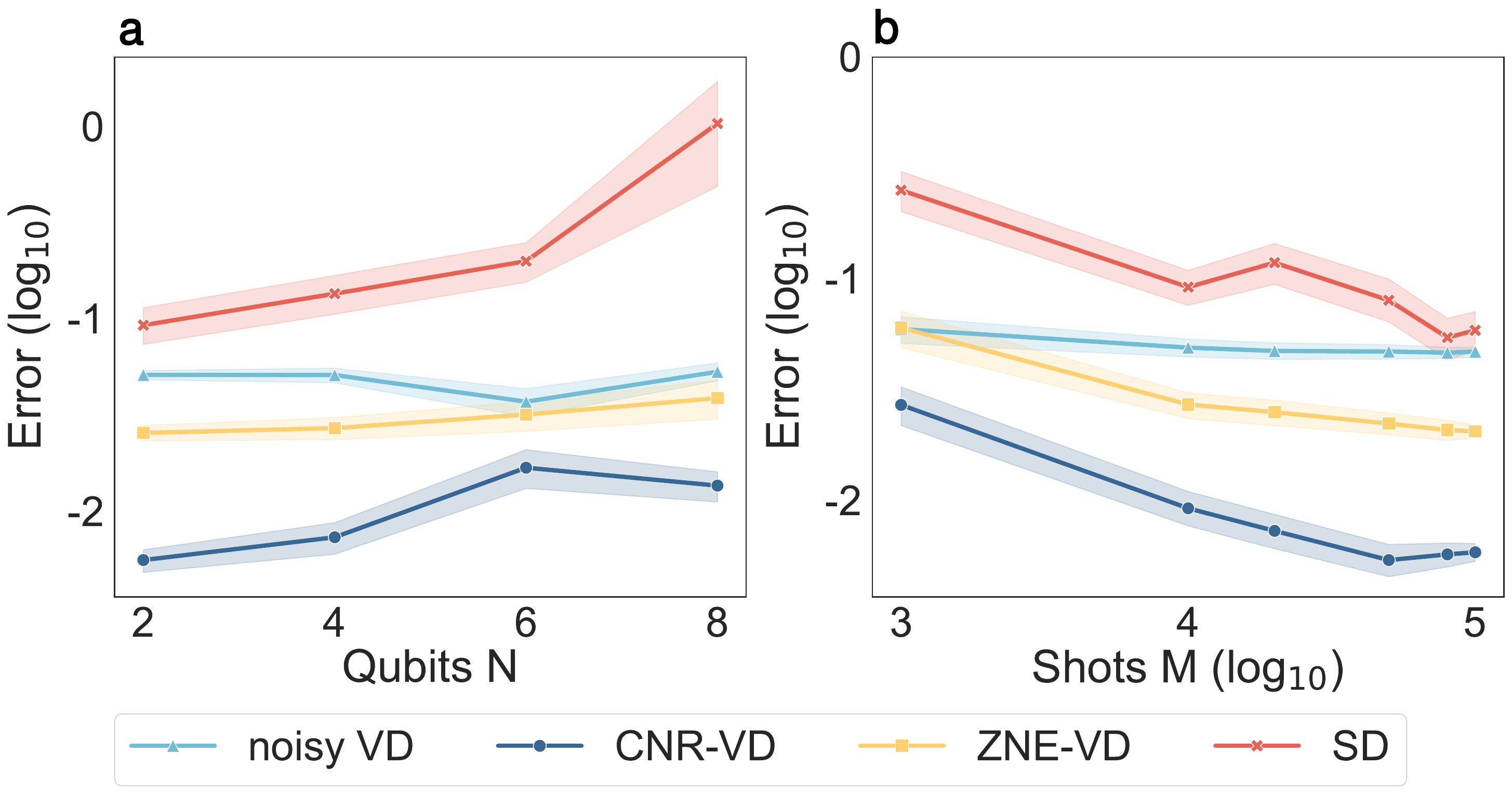}}
    \caption{{\textbf{Accuracy of several VD-based methods for Trotterized random states, with increasing qubit number $N$ and shot number $M$.}} \textbf{a} Performance for noisy VD, CNR-VD, ZNE-VD and SD with increasing qubit numbers. \textbf{b} Performance for four estimators with increasing shot numbers.
    The displayed data represents the average outcome derived from 20 separate experiments, with error bars for standard deviation.}\label{figs-TS-result}
  \end{figure}

In the maintext, we compare the performance on five estimators: CNR-VD, ZNE-VD, SD, noisy VD and unmitigated estimation (Unmit) for $N$-qubit random parameterized state and Pauli observable $\sigma_z^N$, with varying qubit number $N$ and total shot number $M$. The total shot number $M$ contains all the procedures of the estimators.

For CNR-VD, it takes double circuit instances of noisy VD due to the calibration stage. But in this simulation, we fix the observable to be $\sigma_z^N$. Then the coefficient $\hat{O}_\text{VD}(s)$ in Eq.~\ref{eq-est2} can be reused for different noisy state with the same qubit number and noise level. Due to the reusability feature, we have uniformly set the number of measurement shots for calibration in CNR-VD to $10^5$, excluding these from the total measurement shots number used for each noisy state's corresponding VD circuit. Moreover, as we employ RC for CNR-VD with 4 random instances, CNR-VD takes 8 circuit instances and the measuremeant shot number of each circuit instance is $M/8$. Employing the same methodology, one can calculate the measurement shot number for each circuit instance for ZNE-VD as $M/16$.

As discussed in the maintext, the mitigated VD methods including CNR-VD, ZNE-VD and SD introduce additional sampling overhead to mitigate the noise in the VD circuit, leading to poorer results when the total shot number $M$ is low, but CNR-VD and ZNE-VD are still superior to unmitigated estimations and noisy VD. While our simulation results shown in Fig.~\ref{figs-TS-result} favor CNR-VD under the current experimental setup, including the specified composite noise model and measurement resource allocation, we now proceed to elucidate the potential advantages of CNR-VD over ZNE-VD from the perspectives of method design and resource utilization. 

In terms of method design, CNR-VD builds upon the original VD method, refining the outputs of the noisy VD circuit to improve the precision of expectation value estimations. Unlike ZNE-VD, which integrates the Zero-Noise Extrapolation (ZNE) technique with the VD algorithm, CNR-VD avoids the introduction of additional algorithmic errors, as it is specifically tailored to work with the inherent noise in the VD circuit. As for ZNE-VD, while ZNE is a powerful general error mitigation strategy, its integration with VD requires extrapolation to zero noise levels, which can be prone to inaccuracies, especially as the system size increasing. This is due to the complexity of the noise amplification process in ZNE, which becomes more challenging to manage with larger VD circuits. From the perspective of resource utilization, CNR-VD offers a significant advantage. As discussed above, CNR-VD allows for the reuse of coefficients in error-mitigation estimator calculations under the same observable. However, ZNE-VD necessitates multiple executions of the VD algorithm at varying noise levels to calculate error mitigation values for different states. Therefore, under the constraint of an equal total number of measurement shots, CNR-VD can allocate more measurement shots to calculate each expectation value required for VD, compared to ZNE-VD.
This feature not only simplifies the computational process but also optimizes resource use, providing an operational efficiency that is not matched by ZNE-VD.
\begin{figure}[t]
    {\includegraphics[width=1\textwidth]{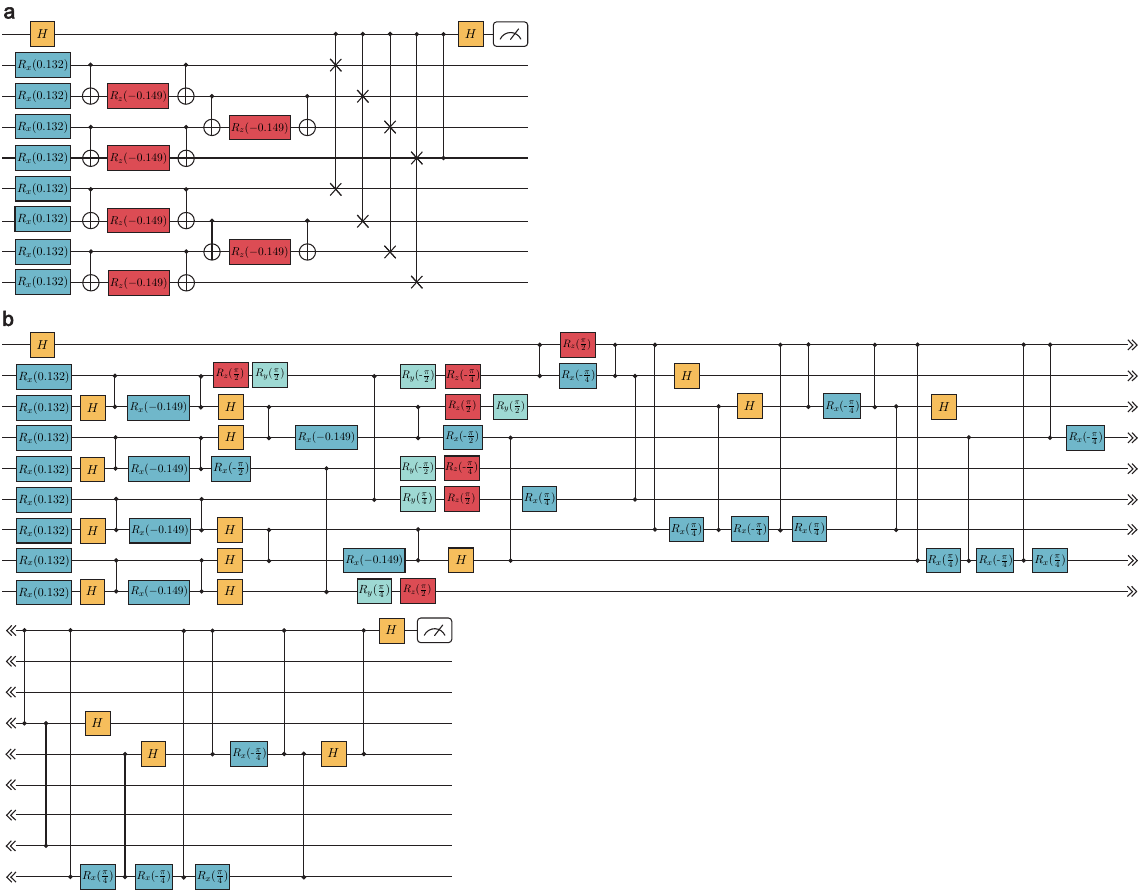}}
    \caption{{\textbf{The illustration for the circuit compilation used in the numerical simulations.} \textbf{a} The circuit used in the VD algorithm for the simulations of the 4-qubit 1D transverse-field Ising model with $\sigma_z^4$. The state preparation circuit is obtained by the first-order trotterized decomposition. \textbf{b} The circuit after compilation with basic gate set of single-qubit and two-qubit gates.}}
    \label{figs-compile}
  \end{figure}

  \begin{figure}[t]
    {\includegraphics[width=.65\textwidth]{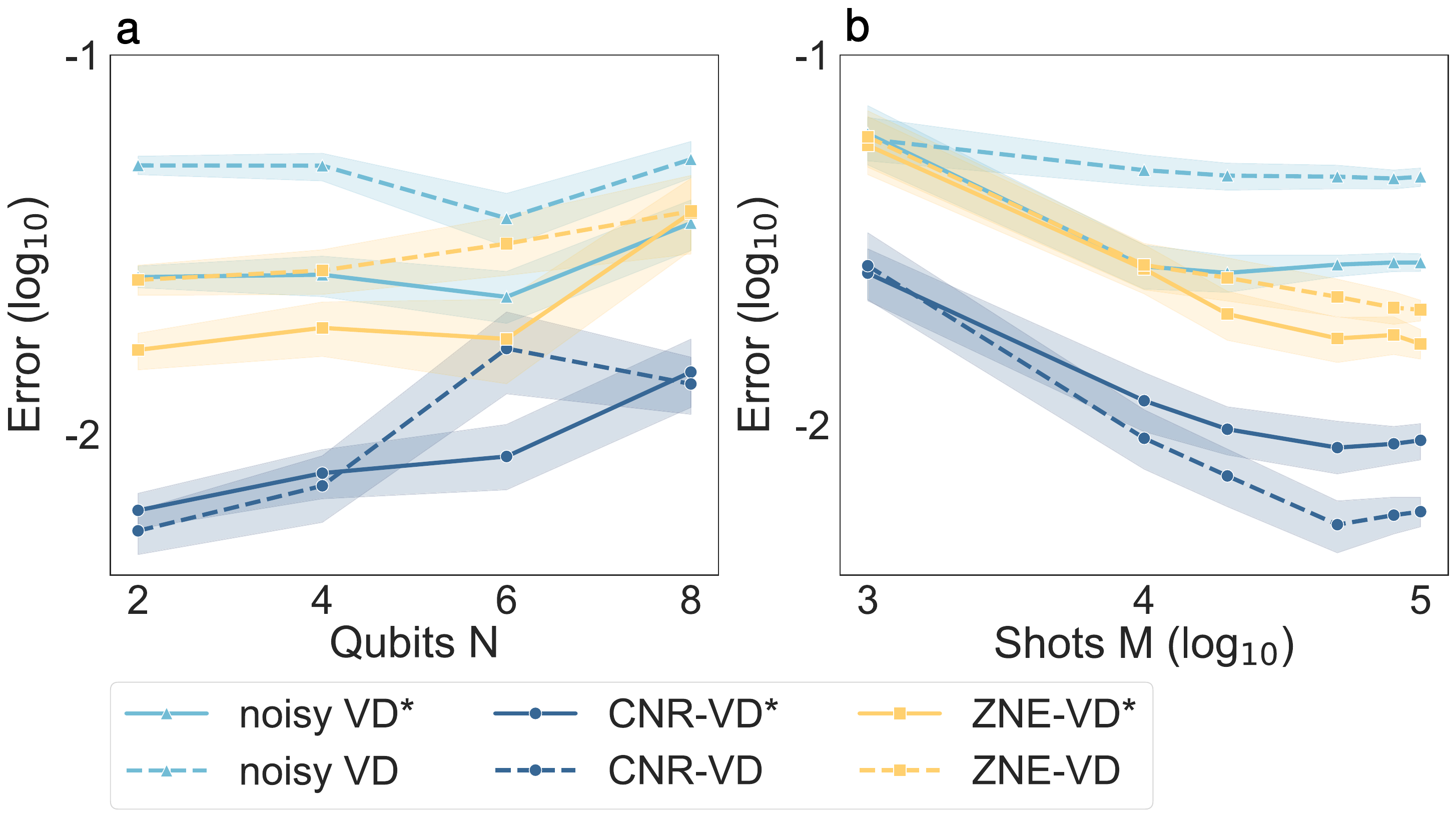}}
    \caption{\textbf{Accuracy of several VD-based methods for random parameterized state, with increasing qubit number $N$ and measurement shot number $M$.} \textbf{a} Performance of six estimators: CNR-VD* (with compilation), CNR-VD, ZNE-VD* (with compilation), ZNE-VD, noisy VD* (with compilation) and noisy VD for random parameterized states, with measurement shots number $2\times 10^4$. \textbf{b} Performance of estimators for 4-qubit random parameterized states. The displayed data represents the average outcome derived from 20 separate experiments for each qubit number and circuit depth, with error bars for standard deviation.}
    \label{figs-compile-result}
  \end{figure}

\subsubsection{Simulations for comparison before and after the compilation process}
Considering the practical application scenarios where circuits are compiled, we expand our investigation to include a comparison of the performance of VD-based algorithms using VD circuits (noisy VD, CNR-VD and ZNE-VD) before and after the compilation process. For the quantum simulation experiments of 1D transverse-field Ising mode, we employ the transpile function of Qiskit~\cite{Qiskit2024qiskit} for circuit compilation when implementing VD-based algorithms, with a compilation level set to 3. Fig.~\ref{figs-compile} illustrates the comparison between the circuit before and after compilation with basic gate set of single-qubit and two-qubit gate controlled-$Z$ gate, which is typical in the superconducting quantum computors. The results demonstrated in Fig.~\ref{figs-compile-result} indicates that the compilation process has a moderate impact on the performance of the VD-based algorithms, yet the performance remains largely consistent. Notably, our CNR-VD demonstrates consistent superiority over other VD-based methods, highlighting its scalability and robust effectiveness in the context of quantum error mitigation. This finding underscores the practical applicability of CNR-VD in real-world quantum computing scenarios.

\bibliographystyle{apsrev4-1}

\bibliography{b}